\pdfoutput=1
\documentclass[preprint,showkeys,secnumarabic,amsfonts,showpacs,amsmath,amssymb]{revtex4}
\usepackage[dvips]{color}
\usepackage{epsfig}
\usepackage{amsmath}
\usepackage{graphicx}

\begin{document}

\title{Cosmic acceleration and phantom crossing in $f(T)$-gravity}

\author{H. Farajollahi$^{1,2}$} \email{hosseinf@guilan.ac.ir} \author{A. Ravanpak$^{1}$}  \author{P.Wu$^{3}$}
\affiliation{Department of Physics,University of Guilan, Rasht, Iran}
\affiliation{School of Physics, University of New South Wales, Sydney, NSW, 2052, Australia}
\affiliation{Center for Nonlinear Science and Department of Physics, Ningbo University,  Ningbo, Zhejiang 315211, China}

\begin{abstract}
In this paper, we propose two new models in $f(T)$ gravity to realize universe acceleration and phantom crossing due to dark torsion in the formalism. The model parameters are constrained and the
observational test are discussed. The best fit results favors an accelerating universe with possible phantom crossing in the near past or future followed respectively by matter and radiation dominated era.
\end{abstract}

\maketitle

\section{Introduction}

In recent cosmological observations such as the Type Ia Supernova (SNe Ia) \cite{Riess}\cite{Perlm}, the cosmic microwave background radiation \cite{Sper}\cite{Sper2} and the large scale structure \cite{Teg}\cite{Eis}, et al., we found that in near past,  the universe is undergoing an accelerating expansion phase. This surprising behavior represents one of the most complex issues in today cosmology. So far, the so called "dark energy" (DE) in our universe, as an exotic energy component with negative pressure dominating the universe is known as the most prominent candidate driving universe to an accelerating expansion phase.

The simple $\Lambda CDM$ model with a constant equation of state (EoS) is known as the simplest model in general agreement with the current experimental data, However, the model suffers from "cosmological constant problem \cite{Car}\cite{Cop}. In addition,  recent observation reveals that the model may endure from an age problem  \cite{Yang1}. An alternative approach thus seems reasonable to find and investigate for the current acceleration of the universe. Observational probes support a small deviations in the EoS parameter by crossing phantum divide line from above to below in the near past. This therefore requires a dynamical description of the parameter \cite{Feng}. The accelerating universe may be interpreted as a failure in our understanding to the gravitational law. Thus we may require an alternative to standard theory of gravity.

In last few years, a variety of cosmological models, such as, quintessence \cite{Cald}, k-essence \cite{Arm},  tachyon \cite{Pad}\cite{Sen},  phantom \cite{Cald2},  quintom \cite{Feng}\cite{Eli};  Chaplygin gas and its generalization \cite{Kam} \cite{Ben}, holographic DE \cite{Cohen}\cite{Li},  agegraphic DE \cite{Wei}\cite{Wei2},  Ricci DE \cite{Gao}  have been investigated.

In addition,  a modification in gravity such as in f(R) \cite{Eliz1}--\cite{San}, or other curvature invariants  \cite{Nojiri1}--\cite{Felice}, by coupling a scalar field to curvature \cite{Nojiri4}\cite{Nojiri3}\cite{Faraj},  a vector field contribution \cite{Zuntz}, or in higher dimensional spacetimes, have widely been studied \cite{Camera}.  Some of these models that display both universe acceleration and phantom crossing are supported by observational probes \cite{Noj}--\cite{Noj4}.

In a remarkably alternative approach to standard general relativity, we use Weitzenböck connection instead of Levi-Civita connection  in the theory where subsequently curvature is replaced by torsion.  In this approach the torsion is formed completely from products of only first derivatives of the tetrad. This theory known as "Teleparallelism", was originally introduced by Einstein in 1928 \cite{Einstein}\cite{Einstein2}. It differs from standard theory of relativity only in "boundary terms" involving total derivatives in the action.  The "Teleparallelism" theory is formulated by gauging external spacetime translation. It employs the Weitzenböck spacetime which is characterized by the metricity condition and vanishing of the curvature tensor. The approach contains a variety of distinctive manifestation both from physical and geometrical aspects \cite{Hoff}--\cite{Yang}.

There are models based on the teleparallel equivalent of general relativity (TEGR) that presented as an alternative to inflationary models with no need to inflaton field \cite{Fer2}\cite{Pop}. Also there are DE models constructed from teleparallelism that explain current acceleration of the universe without using DE component \cite{Wu}--\cite{Bamba} In these models dark torsion (DT) is responsible for the acceleration of the universe, and the field equations are always 2nd order equations.  Among other advantages, this characteristic makes these theories simpler than the resulting in f(R) theories. Note also that in generalized theories of TEGR, where the Lagrangians are algebraic functions of the usual teleparallel Lagrangian, both actions and field equations are not invariant under local Lorentz transformations \cite{Li}. However, the usual teleparallel theory equivalent to general relativity, is just a special case of the so-called $f(T)$ gravity where the Lagrangian of teleparallel gravity, is substituted by $T$ The $f(T)$ gravity provides an alternative to conventional dark energy in general relativistic cosmology to explain the acceleration of the universe.

Note that the main requirement of teleparallelism and $f(T)$ gravity is that there
exist a class of frames where the spin connection vanishes, but torsion does not \cite{Sotiriou}. In fact, the torsion tensor
is formed solely from products of first derivatives of the tetrad and the spacetime torsion manifests itself as a generator of gravitational repulsion. This enables the theory to produce universe acceleration (for example, see \cite{Shie}--\cite{Dent})

The original motivation for considering $f(T)$ gravity is to account for the late time cosmic acceleration without the need for a cosmological constant and/or dark energy. Therefore, it makes sense to restrict ourselves to specific models that exhibit this property. In this manuscript, we follow the authors in \cite{Wu} and introduce two new forms of $f(T)$ models to investigate the universe acceleration and phantom crossing by best fitting the models with the observational data.

\section{GENERAL CONSIDERATIONS}

Teleparallelism uses a vierbein
field $e_i(x^\mu)$, $i = 0, 1, 2, 3$, as a dynamical object which is an orthonormal basis for the tangent space at each point $x^\mu$ of the manifold: $e_i . e_j=\eta_{ij}$ , where $\eta_{ij}=diag(1,-1,-1,-1)$. The vector $e_i$ can be described by its components $e^\mu_i$, $\mu=0,1,2,3$ in a coordinate basis; i.e. $e_i=e^\mu_i\partial_\mu$.
Note that Latin indices refer to the tangent space,
while Greek indices label coordinates on the manifold.
The metric tensor can be obtained from the dual vierbein as $g_{\mu\nu}(x)=\eta_{ij} e^i_\mu(x)e^j_\nu(x)$. Though general relativity uses the torsionless Levi-Civita connection, in Teleparallelism we apply the torsionless Levi-Civita connection,
Teleparallelism uses the curvatureless Weitzenb\"{o}ck connection, whose non-null torsion is
\begin{equation}\label{torsion}
    T^\lambda_{\mu\nu}=\hat{\Gamma}^\lambda_{\nu\mu}-\hat{\Gamma}^\lambda_{\mu\nu}=e^\lambda_i(\partial_\mu e^i_\nu - \partial_\nu e^i_\mu).
\end{equation}
The above equation contains all the information about the gravitational field. The TEGR Lagrangian is constructed from torsion equation, (\ref{torsion}), and the dynamical equations for vierbein are corresponding to the Einstein field equations for the metric

The
teleparallel Lagrangian is
\begin{equation}\label{lagrangian}
    T={S_\rho}^{\mu\nu}{T^\rho}_{\mu\nu},
\end{equation}
where
\begin{equation}\label{s}
    {S_\rho}^{\mu\nu}=\frac{1}{2}({K^{\mu\nu}}_\rho+\delta^\mu_\rho {T^{\theta\nu}}_\theta-\delta^\nu_\rho {T^{\theta\mu}}_\theta)
\end{equation}
and ${K^{\mu\nu}}_\rho$ is the contorsion tensor
\begin{equation}\label{contorsion}
    {K^{\mu\nu}}_\rho=\frac{-1}{2}({T^{\mu\nu}}_\rho-{T^{\nu\mu}}_\rho-{T_\rho}^{\mu\nu}),
\end{equation}
that is equal to the difference between Weitzenb\"{o}ck and Levi-Civita connections.

In this work we start with the following action where Lagrangian density sums over $T$ and $f(T )$. Thus the action reads
\begin{equation}\label{action}
    I = \frac{1}{16\pi G}\int{d^4xe(T+f(T))},
\end{equation}
where $e=det(e^i_\mu)=\sqrt{-g}$. The action only with $T$ corresponds to TEGR. In the presence of  matter field coupled to the metric,  variation of action with respect to the vierbein yields the following equations \cite{Ben2}
\begin{eqnarray}
  e^{-1}\partial_\mu(e{S_i}^{\mu\nu})(1+f^{'}(T))-e_i^\lambda {T^\rho}_{\mu\lambda}{S_\rho}^{\nu\mu}(1+f^{'}(T))\nonumber \\+{S_i}^{\mu\nu}\partial_\mu(T)f^{''}(T)+\frac{1}{4}e^\nu_i(T+f(T))=4\pi G{e_i}^\rho {T_\rho}^\nu, \label{equations}
\end{eqnarray}
where prime denotes differentiation with respect to $T$,
${S_i}^{\mu\nu}={e_i}^\rho {S_\rho}^{\mu\nu}$ and $T_{\mu\nu}$ is the matter energy-momentum
tensor.

\section{COSMOLOGICAL SOLUTION AND OBSERVATIONAL CONSTRAINTS}
We assume a flat homogeneous and isotropic FRW universe such that
\begin{equation}\label{metric}
    e^i_\mu = diag(1, a(t), a(t), a(t)),
\end{equation}
where $a(t)$ is the cosmological scale factor. By using (\ref{torsion}), (\ref{lagrangian}), (\ref{s}) and (\ref{contorsion}) we obtain
\begin{equation}\label{lt}
    T=-6H^2,
\end{equation}
where $H=\frac{\dot a}{a}$, being the Hubble parameter. The substitution of the vierbein (\ref{metric}) in (\ref{equations})
for $i=0=\nu$ yields
\begin{equation}\label{friedmann}
    T(1+2f^{'}(T))-f(T)=-16\pi G\rho.
\end{equation}
Besides, the equation $i=1=\nu$ is
\begin{eqnarray}\label{acceleration}
    \frac{2}{3}\sqrt{-6T}\dot{T}f^{''}(T)+f^{'}(T)[2T+\frac{\dot{T}}{3}\sqrt{\frac{-6}{T}}]-f(T) \nonumber \\
    +\frac{\dot{T}}{3}\sqrt{\frac{-6}{T}}+T=16\pi Gp.
\end{eqnarray}
In equations (\ref{friedmann}) and (\ref{acceleration}), $\rho$ and $p$ are the dark matter energy density and pressure, respectively. It can be easily derived that they accomplish the conservation equation
\begin{equation}\label{conservation}
    \dot{\rho}+\sqrt{\frac{-3T}{2}}(\rho+p)=0.
\end{equation}
Here we assume that there are both matter and radiation components in the Universe, thus $\rho = \rho_m + \rho_r$ and $p = \frac{1}{3}\rho_r$. If we rewrite the modified Friedmann equation (\ref{friedmann}) in the standard form as that in general relativity, we can define a torsion energy density, expressed as
\begin{equation}\label{rhoeff}
    \rho_{T} = \frac{1}{16\pi G}(2Tf'(T)-f(T)).
\end{equation}
An assumption that $\rho_{T}$ satisfies the torsion EoS, $p_{T} = \omega_{T}\rho_{T}$ and by using (\ref{acceleration}) to define the torsion pressure $p_{T}$, one can readily obtains the EoS parameter correspond to the DT as
\begin{equation}\label{omegaeff}
    \omega_{T}=-1-\frac{A_1}{B_1},
\end{equation}
where
\begin{equation}
    A_1=(f'(T)+2Tf''(T))(1+2f'(T)-\frac{f(T)}{T}+\frac{\Omega_r}{3})
\end{equation}
and
\begin{equation}
    B_1=(1+f'(T)+2Tf''(T))(\frac{f(T)}{T}-2f'(T)).
\end{equation}
It is clear from (\ref{omegaeff}) that the necessary condition for crossing the phantom divide line ($\omega_{T}=-1$) is either
\begin{equation}\label{condition1}
    f'(T)+2Tf''(T)=0,
\end{equation}
or
\begin{equation}\label{condition2}
    \Omega_r=-3(1+2f'(T)-\frac{f(T)}{T}).
\end{equation}
In addition, a dynamical EoS parameter also needs to satisfy the condition:
\begin{equation}\label{othercon}
    \frac{d}{dt}(\rho_T+p_T)\neq0,
\end{equation}
when $\omega_T\rightarrow-1$. Thus, one requires that the second term in the equation (\ref{omegaeff}) changes sign in order for the parameter to exhibit a dynamical behavior.

In the following, we are interested to examine the accelerated expansion universe and phantom crossing driven by torsion. For this
we try with two $f(T)$ models. We compare our both models with the $\Lambda$CDM model and also the SNe Ia observational data, combined with the information coming from the BAO (Baryon Acoustic Oscillation) and the CMB shift parameter.

{\bf First Model:}

We consider the $f(T)$ function as
\begin{equation}\label{f}
    f(T)=\alpha(-T)^n\sinh(\frac{{T}_0}{T}),
\end{equation}
where $\alpha$ and $n$ are two model parameters. This model is different from the one introduced in \cite{Wu} by multiplying $f(T)$ by $\sinh(\frac{{T}_0}{T})$ instead of $\tanh(\frac{{T}_0}{T})$. Substituting (\ref{f}) into the modified Friedmann equation (\ref{friedmann}), we have
\begin{equation}\label{alpha}
    \alpha=\frac{1-{\Omega_m}_0-{\Omega_r}_0}{(6{H_0}^2)^{n-1}[(2n-1)\sinh(1)-2\cosh(1)]}.
\end{equation}
Here ${\Omega_m}_0$ and ${\Omega_r}_0$ are the present dimensionless density parameters of matter and radiation,
respectively. From $f(T)$ given in (\ref{f}), one can rewrite the DT energy density for this model as
\begin{equation}\label{rhoeff1}
    \rho_{T} = \frac{\alpha(-T)^n}{16\pi G}[(2n-1)\sinh(\frac{{T}_0}{T})-2\frac{{T}_0}{T}\cosh(\frac{{T}_0}{T})].
\end{equation}
To satisfy energy condition, we must have $\rho_{T}\geq0$. However, from numerical computation we obtain that for $n>\frac{3}{2}$ the function $\rho_T<0$ in some region. Therefore, in order to fulfill the energy condition requirement, the first constraint for $n$ is $n\leq\frac{3}{2}$.

Also, to satisfy the equations (\ref{condition1}) or (\ref{condition2}), the necessary condition to cross the phantom divide line, $-1$, is
\begin{equation}\label{omegacross11}
   \tanh\left(\frac{T_0}{T}\right)=\frac{(4n-3)T_0T}{nT^2(2n-1)+2T_0^2}\,\cdot
\end{equation}
or
\begin{eqnarray}\label{omegacross2}
   \Omega_r=-3+3\alpha(-T)^{n-2}((1-2n)T\sinh\left(\frac{T_0}{T}\right)\nonumber \\
   +2T_0\cosh\left(\frac{T_0}{T}\right)).
\end{eqnarray}
A dynamical dark torsion EoS parameter require to satisfy one of the above conditions in addition to the constraint (16) in our model. The given $f(T)$ model satisfies the condition $f(T)/T \rightarrow 0 $ and therefore is consistent with the primordial nucleosynthesis and cosmic microwave background constraints \cite{Wu2}. Furthermore, for $\alpha \rightarrow 0$, one regains the usual general relativity for an spatially flat FRW cosmology.

{\bf Second Model:}

We consider $f(T)$ as
\begin{equation}\label{f2}
    f(T)=\alpha(-T)^n\cosh(\frac{p{T}_0}{T})(1-\exp(\frac{p{T}_0}{T})),
\end{equation}
with three model parameters $\alpha$, $n$ and $p$. This model is also different from the one introduced in \cite{Wu} by multiplying $f(T)$ by $\cosh(\frac{p{T}_0}{T})$. By using (\ref{friedmann}), we can rewrite the parameter $\alpha$ in terms of other cosmological parameters as
\begin{equation}\label{alpha2}
    \alpha=\frac{1-{\Omega_m}_0-{\Omega_r}_0}{(6{H_0}^2)^{n-1}C_1}
\end{equation}
where
\begin{eqnarray}
    C_1=2\cosh(p)[pe^p+n(1-e^p)]+(e^p-1)\times\nonumber \\
    \left(\cosh(p)+2p\sinh(p)\right).
\end{eqnarray}
From $f(T)$ given in (\ref{f2}), one can rewrite the DT energy density for this model as
\begin{eqnarray}\label{rhoeff2}
    \rho_{T} &=& \frac{\alpha(-T)^n}{16\pi G}[2\cosh(\frac{p{T}_0}{T})[\frac{p{T}_0}{T}e^{\frac{p{T}_0}{T}}+n(1-e^{\frac{p{T}_0}{T}})]\nonumber\\
    &+&(e^{\frac{p{T}_0}{T}}-1)[\cosh(\frac{p{T}_0}{T})-2\frac{p{T}_0}{T}\sinh(\frac{p{T}_0}{T})]].
\end{eqnarray}
Similar to the first model, for energy condition, $\rho_{T}\geq0$, we need the constraint $n\leq1.44$ rounding up to 2 decimal. In order for $\omega_T$ crosses the phantom line, $-1$, the conditions (\ref{condition1}) and (\ref{condition2}) for this model become
\begin{equation}\label{omegacross33}
    \tanh\left(\frac{T_0}{T}\right)=\frac{\exp\left(\frac{pT_0}{T}\right)A_2}{B_2}
\end{equation}
where
\begin{eqnarray}\label{omegacross33}
    A_2=(nT^2(1-2n)-pT_0T(3-4n)-4p^2T_0^2)\nonumber \\
    -(nT^2(1-2n)-2p^2T_0^2)
\end{eqnarray}
and
\begin{equation}\label{omegacross33}
    B_2=pT_0T(4n-3)+\exp\left(\frac{pT_0}{T}\right)(pT_0T(3-4n)+4p^2T_0^2)
\end{equation}
and
\begin{eqnarray}\label{omegacross4}
    \Omega_r=-3+6\alpha pT_0(-T)^{n-2}\left(1-\exp\left(\frac{2pT_0}{T}\right)\right)\nonumber\\
    +3\alpha(2n-1)(-T)^{n-1}\cosh\left(\frac{pT_0}{T}\right)\left(1-\exp\left(\frac{pT_0}{T}\right)\right).
\end{eqnarray}
Again, a dynamical dark torsion EoS parameter require to satisfy one of the above conditions in addition to constraint (16) for this model. Similar to the first model, the given $f(T)$ satisfies the condition $f(T)/T \rightarrow 0$ and therefore is consistent with the primordial nucleosynthesis and cosmic microwave background constraints. Note also that for $ \alpha\rightarrow 0$ or $p \rightarrow 0$, the usual general relativity is retrieved.

\section{Cosmological constraints and tests }

In the following we first best fit the model parameters with observational data for luminosity distance of SNe Ia , the baryonic acoustic oscillation (BAO) distance ratio and the cosmic microwave background (CMB) radiation,  using ?2 method. We then analyse the behavior of dark torsion EoS parameter and total EoS parameter predicted by model  for the best fitted parameters.

\subsection{Cosmological model constraints}

From observations, the Luminosity distance quantity, $D_L(z)$, determines DE density. In the theoretical model with the $H(z)$ obtained from numerical computation, we obtain  \cite{Wu}
\begin{equation}\label{dl}
D_{L}(z)\equiv(1+z)\int_0^z{\frac{dz'}{E(z')}},
\end{equation}
where $E(z)\equiv H(z)/H_0$. The difference between the absolute and apparent luminosity of a distance object is called distance modulus, $\mu(z)$, and given by, $\mu(z) = 5\log_{10}D_L(z) - \mu_0$ where $\mu_0 = 5 log_{10}h + 42.38$ and $h = (H_0/100)$km/s/Mpc.

Now, we obtain the constraints on the model parameters utilizing recent observational data, including SNe Ia which consists of 557 data points and belongs to the Union sample \cite{Aman}, BAO distance ratio and CMB  radiation. To constrain the parameters in the models from the SNe Ia, to best fit the observational data, one employs the $\chi^2$ value
\begin{equation}\label{chi2}
    \chi^2_{Sne}=\sum_{i=1}^{557}\frac{[\mu_i^{the}(z_i) - \mu_i^{obs}(z_i)]^2}{\sigma_i^2},
\end{equation}
where summation is over the cosmological data points. In (\ref{chi2}),  $\mu_i^{the}$ and $\mu_i^{obs}$ are the distance modulus obtained from model and  observation, respectively. Also, $\sigma_i$ is the estimated error of the $\mu_i^{obs}$ where obtained from observation.

From joint analysis of the 2dF Galaxy Redshift Survey and SDSS data \cite{Reid}\cite{Percival}, by using BAO data,  the BAO distance ratio at $z = 0.20$ and $z = 0.35$  is obtained
\begin{equation}\label{drbao}
   \frac{D_V(z=0.35)}{D_V(z=0.20)}=1.736\pm0.065,
\end{equation}
is a model independent measure with $D_V (z)$ given by
\begin{equation}\label{bao}
    D_V(z_{BAO})=[\frac{z_{BAO}}{H(z_{BAO})}(\int_0^{z_{BAO}}\frac{dz}{H(z)})^2]^{1/3}.
\end{equation}
Therefore, one can obtain the constraint from BAO by performing the following $\chi^2$ statistics
\begin{equation}\label{chibao}
    \chi^2_{BAO}=\frac{[(D_V(z=0.35)/D_V(z=0.20))-1.736]^2}{0.065^2}\cdot
\end{equation}
We finally use the CMB data to constrain our model parameters. From CMB data, the CMB shift parameter R \cite{Wang}\cite{Bond}, contains the major observational information. Thus we use it to constrain the model parameters by minimizing
\begin{equation}\label{chicmb}
    \chi^2_{CMB}=\frac{[R-R_{obs}]^2}{\sigma_R^2},
\end{equation}
where $R_{obs} = 1.725\pm0.018$ \cite{Kom}, is given by the WMAP7 data. Its corresponding theoretical value is defined as
\begin{equation}\label{r}
    R\equiv\Omega_{m0}^{1/2}\int_0^{z_{CMB}}\frac{dz'}{E(z')},
\end{equation}
with $z_{CMB} = 1091.3$. The constraints from a combination of Sne Ia, BAO and CMB can be obtained by minimizing $\chi^2_{Sne}+\chi^2_{BAO}+\chi^2_{CMB}$.

\begin{figure}
\centering
\includegraphics[width=0.4\textwidth]{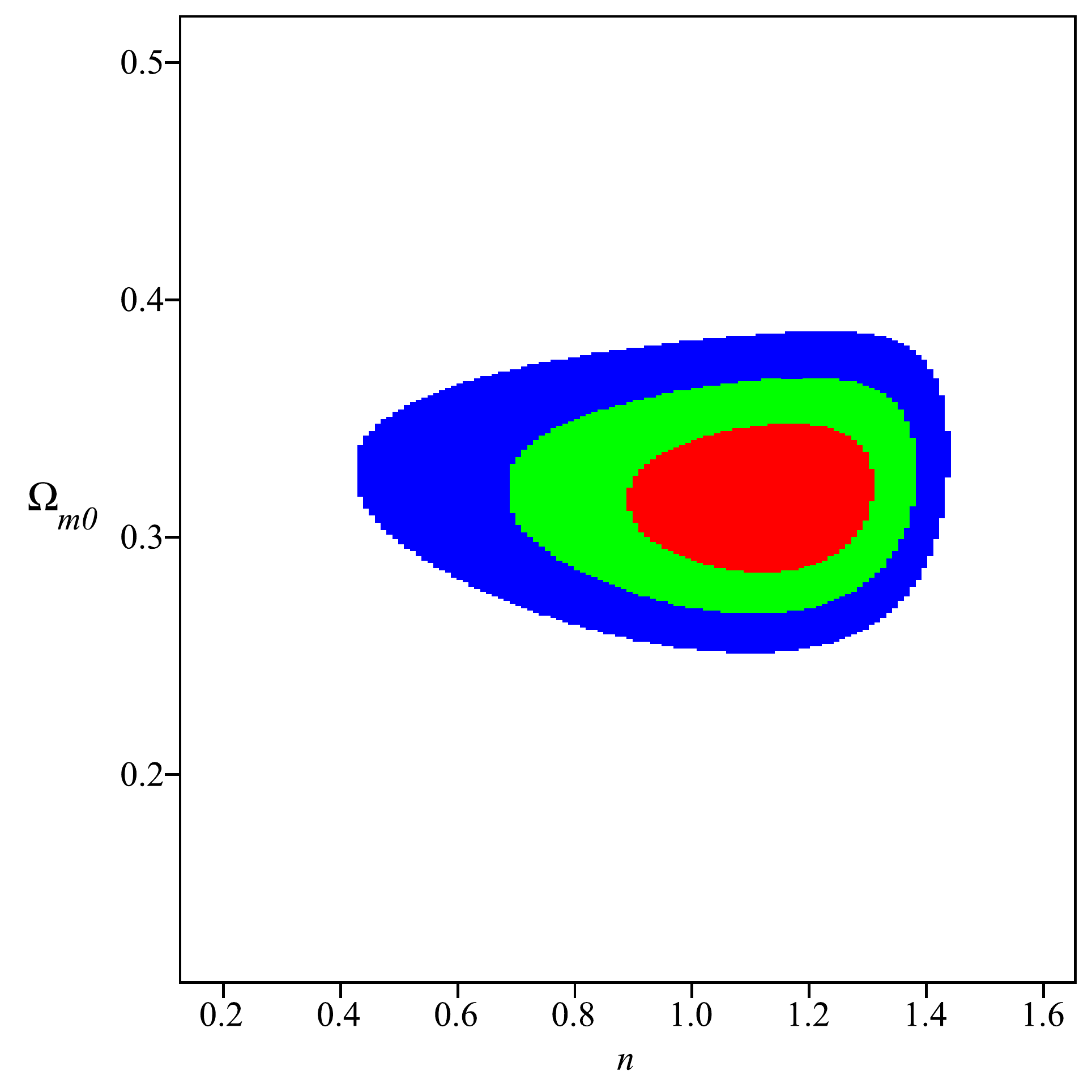}
\caption{The constraint on $\Omega_{m0}$ and $n$ at the 68.3\%, 95.4\% and 99.7\% confidence levels from Sne Ia + BAO + CMB for the first model.}\label{fig5}
\end{figure}

\begin{figure}
\centering
\includegraphics[width=0.32\textwidth]{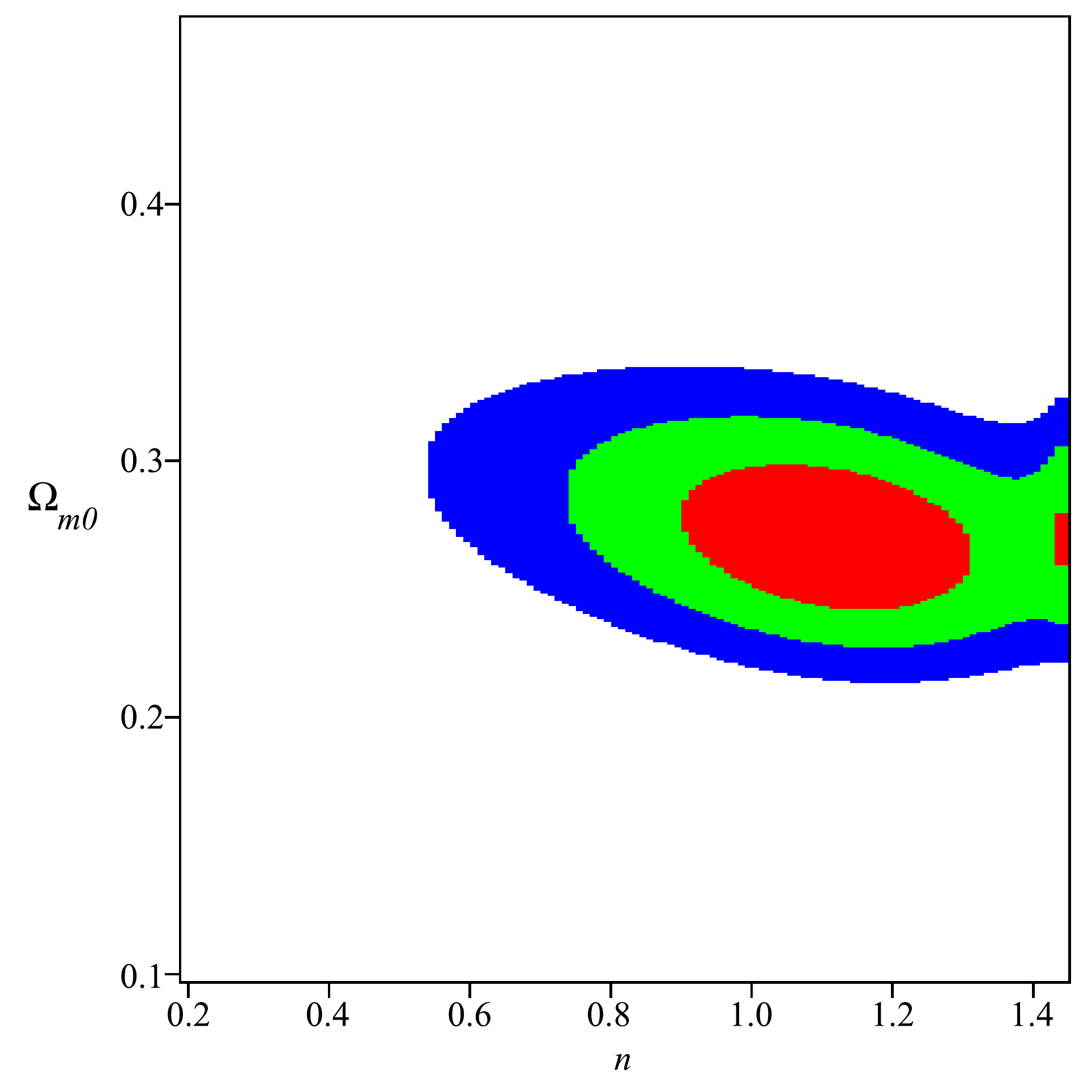}
\includegraphics[width=0.32\textwidth]{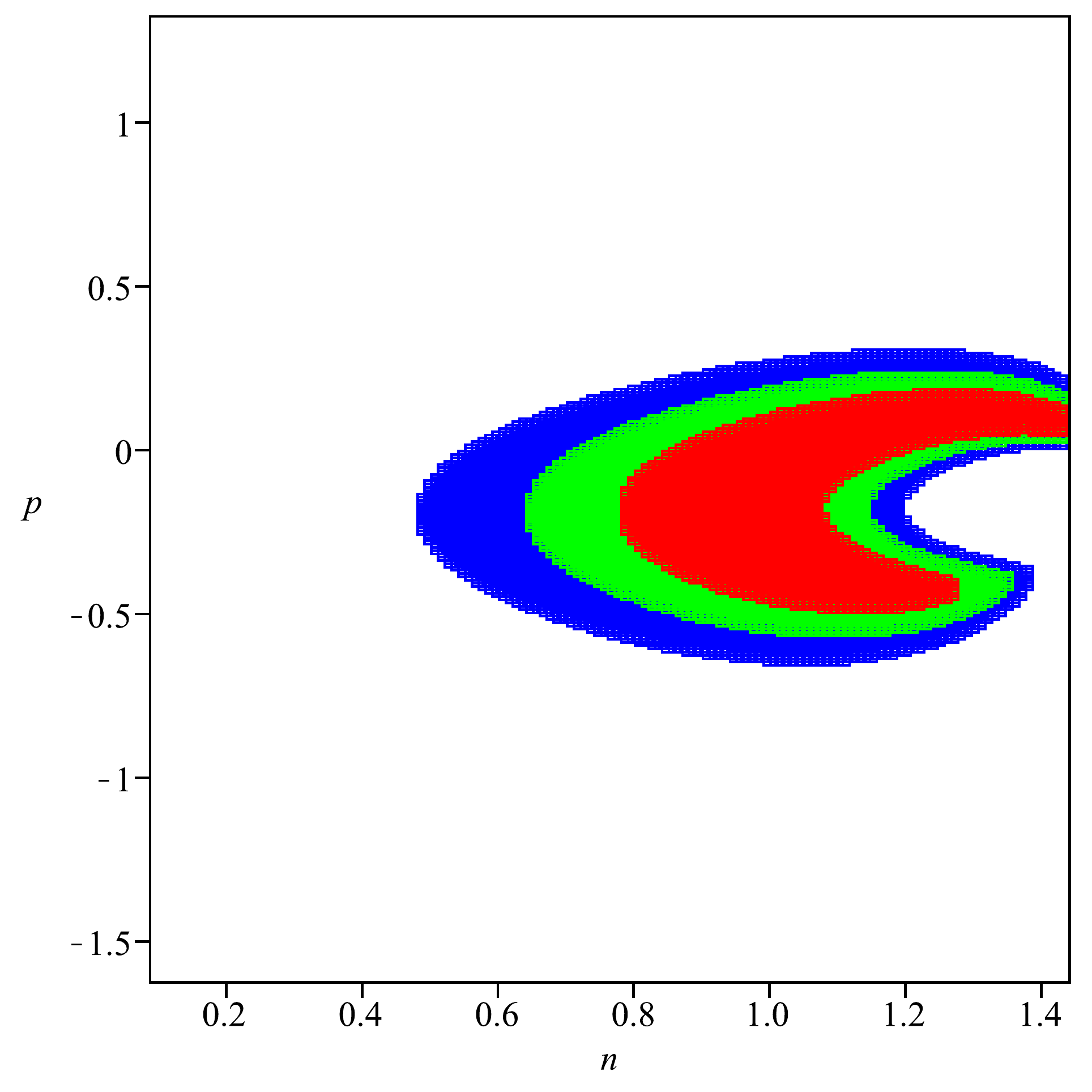}
\includegraphics[width=0.32\textwidth]{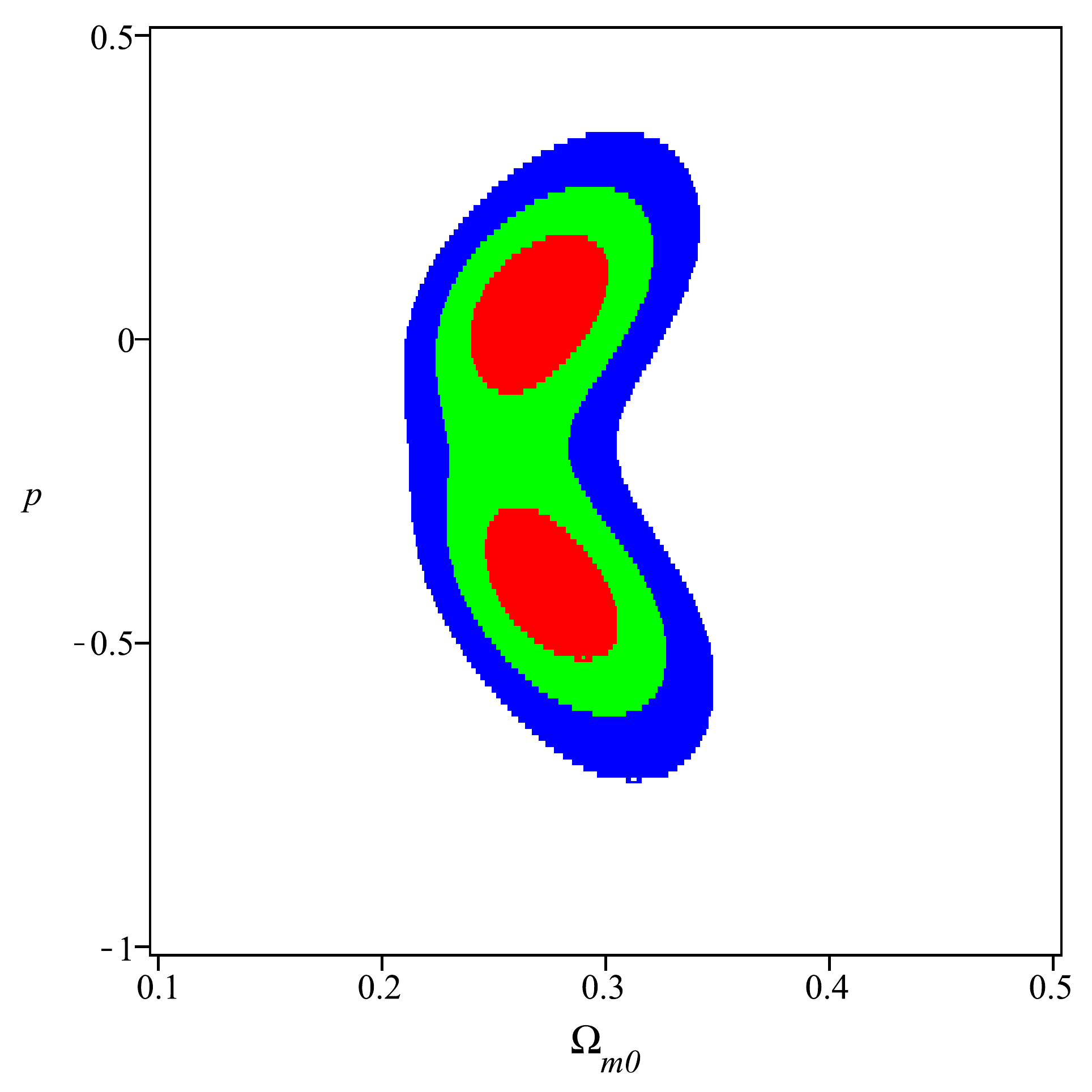}
\caption{The constraint on $\Omega_{m0}$, $n$ and $p$ at the 68.3\%, 95.4\% and 99.7\% confidence levels from Sne Ia + BAO + CMB for the second model.
From top to bottom: ($\Omega_{m0}$-$n$) plane, ($p$-$n$) plane and ($p$-$\Omega_{m0}$) plane.}\label{fig6}
\end{figure}

For the first model, from numerical computation, we find that the best fit values occur at
$\Omega_{m0} = 0.316$ and $n = 1.14$ with $\chi^2_{min}=560.86061$. The contour diagrams at the $68.3\%$, $95.4\%$ and $99.7\%$ confidence levels are given in FIG. (\ref{fig5}). From the graph, one conclude that with $68.3\%$, $95.4\%$ and $99.7\%$ confidence level the true values for both $n$ and  $\Omega_{m0}$ lie within the red, green and blue  contours, respectively. For the second model, the best fit value of model parameters are $\Omega_{m0} = 0.269$, $p = 0.05$ and $n = 1.12$ with $\chi^2_{min}=544.20749$. FIG. (\ref{fig6}) shows the constraints on the model parameters at the $68.3\%$, $95.4\%$ and $99.7\%$ confidence levels.

\begin{figure}
\centering
\includegraphics[width=0.4\textwidth]{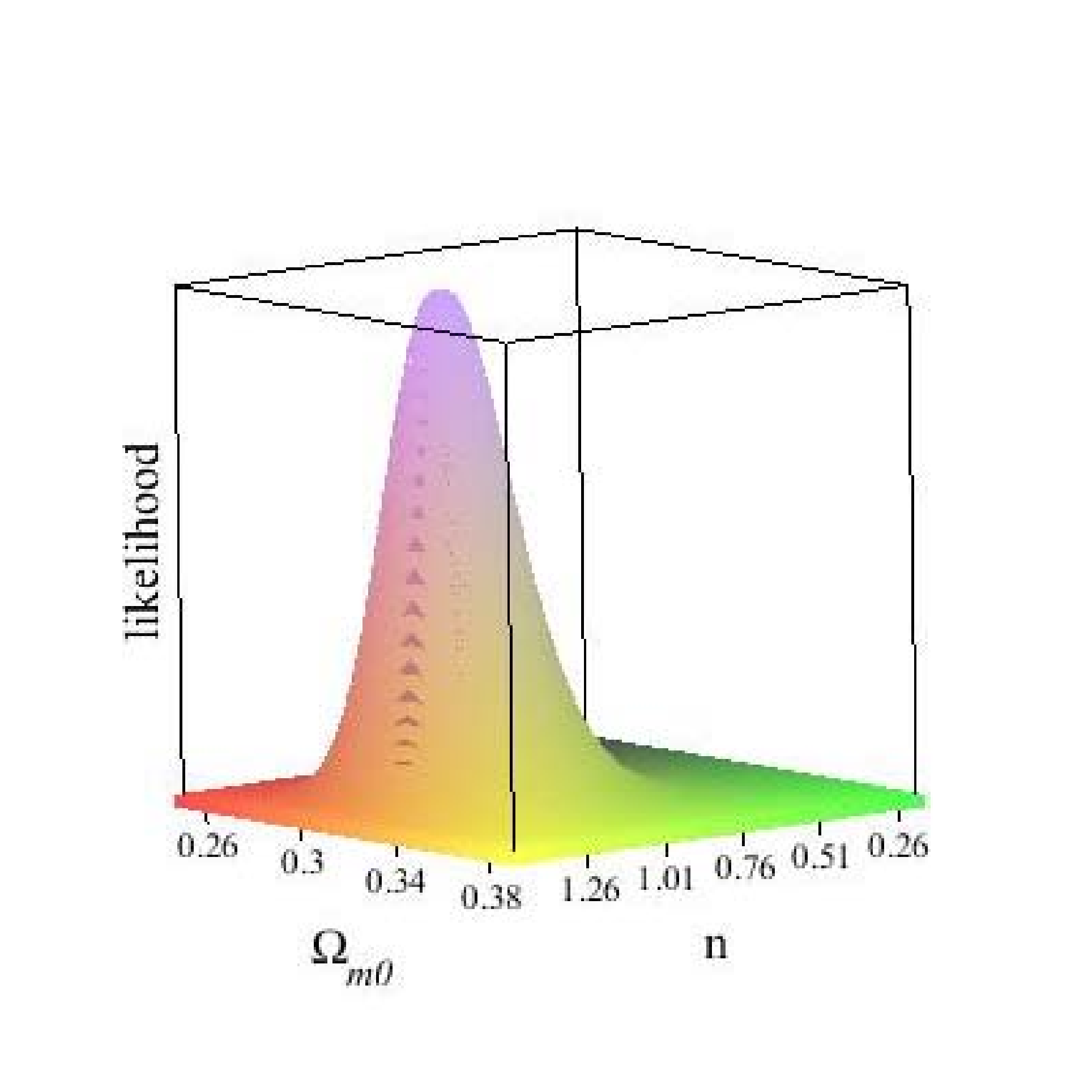}
\caption{Likelihood for the model parameters in the first model.}\label{fig9}
\end{figure}

\begin{figure}
\centering
\includegraphics[width=0.32\textwidth]{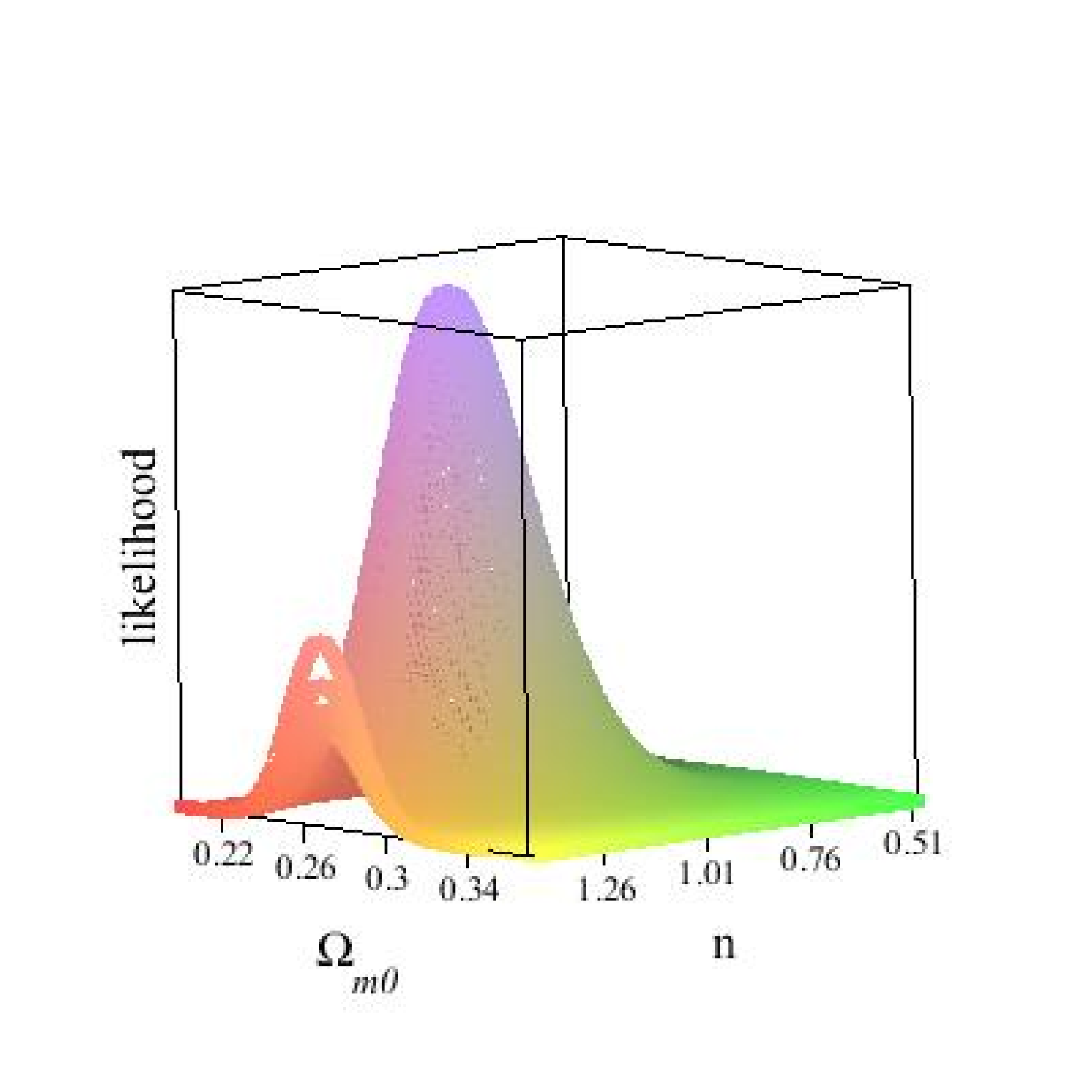}
\includegraphics[width=0.32\textwidth]{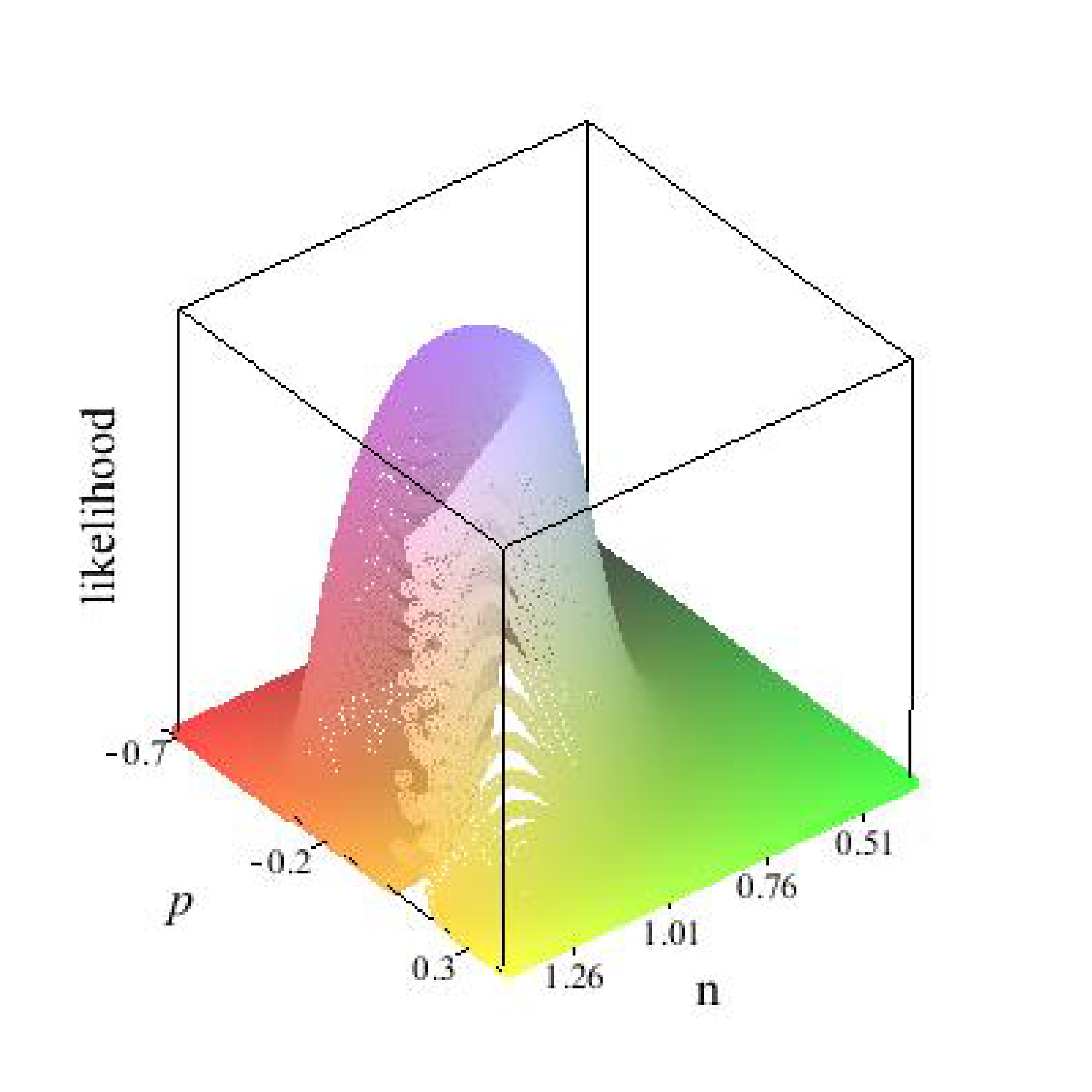}
\includegraphics[width=0.32\textwidth]{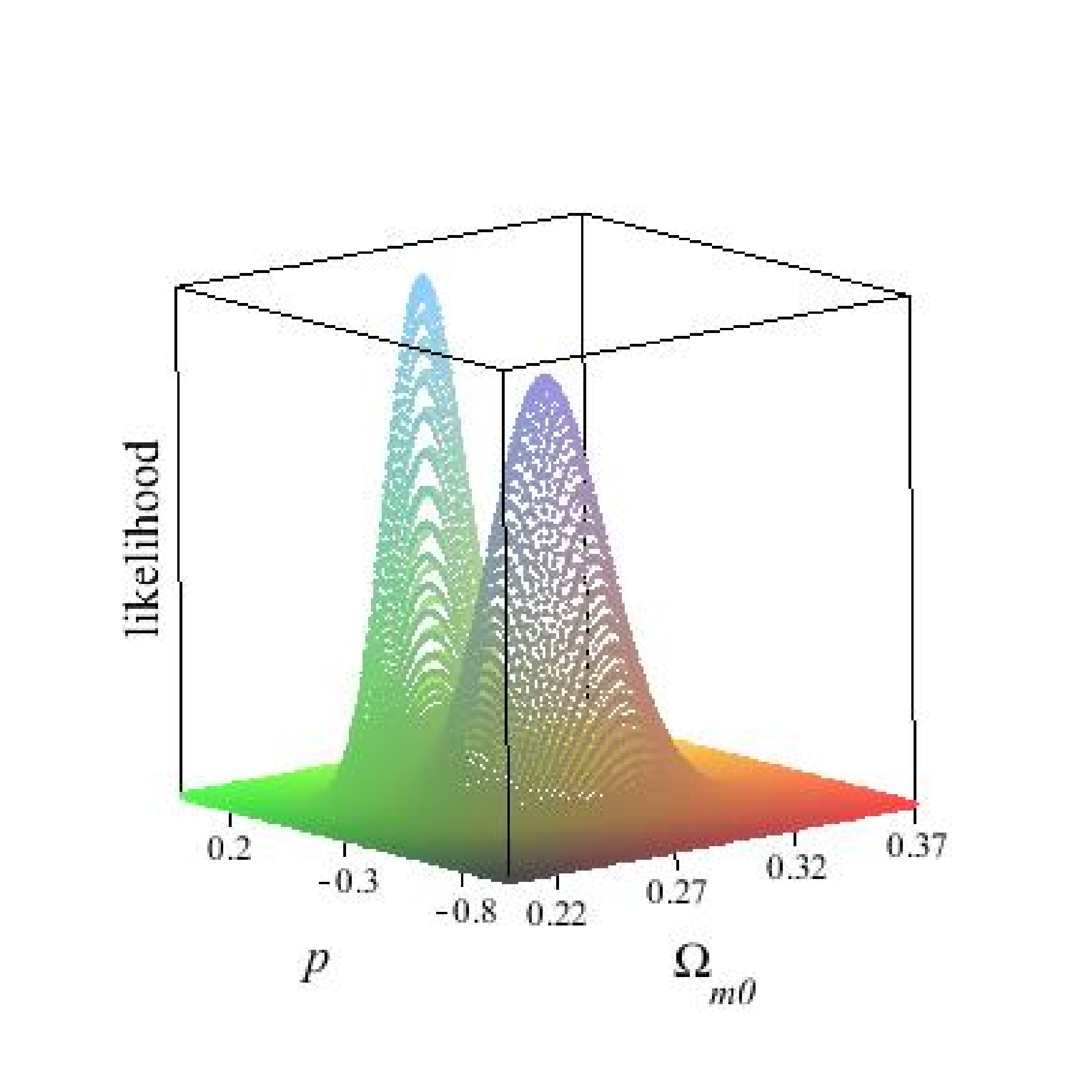}
\caption{Likelihood for the model parameters in the second model.}\label{fig10}
\end{figure}

Alternatively, we can plot the likelihood for the pair model parameters in both cases(FIGS (\ref{fig9}) and (\ref{fig10})).
Obviously in the first model since we only have one pair of parameters, there is only one likelihood plot for it.
In the second model for the three parameters we have likelihood for three pairs of parameters.
In FIG. (\ref{fig10}), the bottom graph (the likelihood for the parameters  $p$ and $\Omega_{m0}$), one observes that there are two peaks, with the best fitted parameters are in the highest peak.

\

\begin{figure}
\centering
\includegraphics[width=0.48\textwidth]{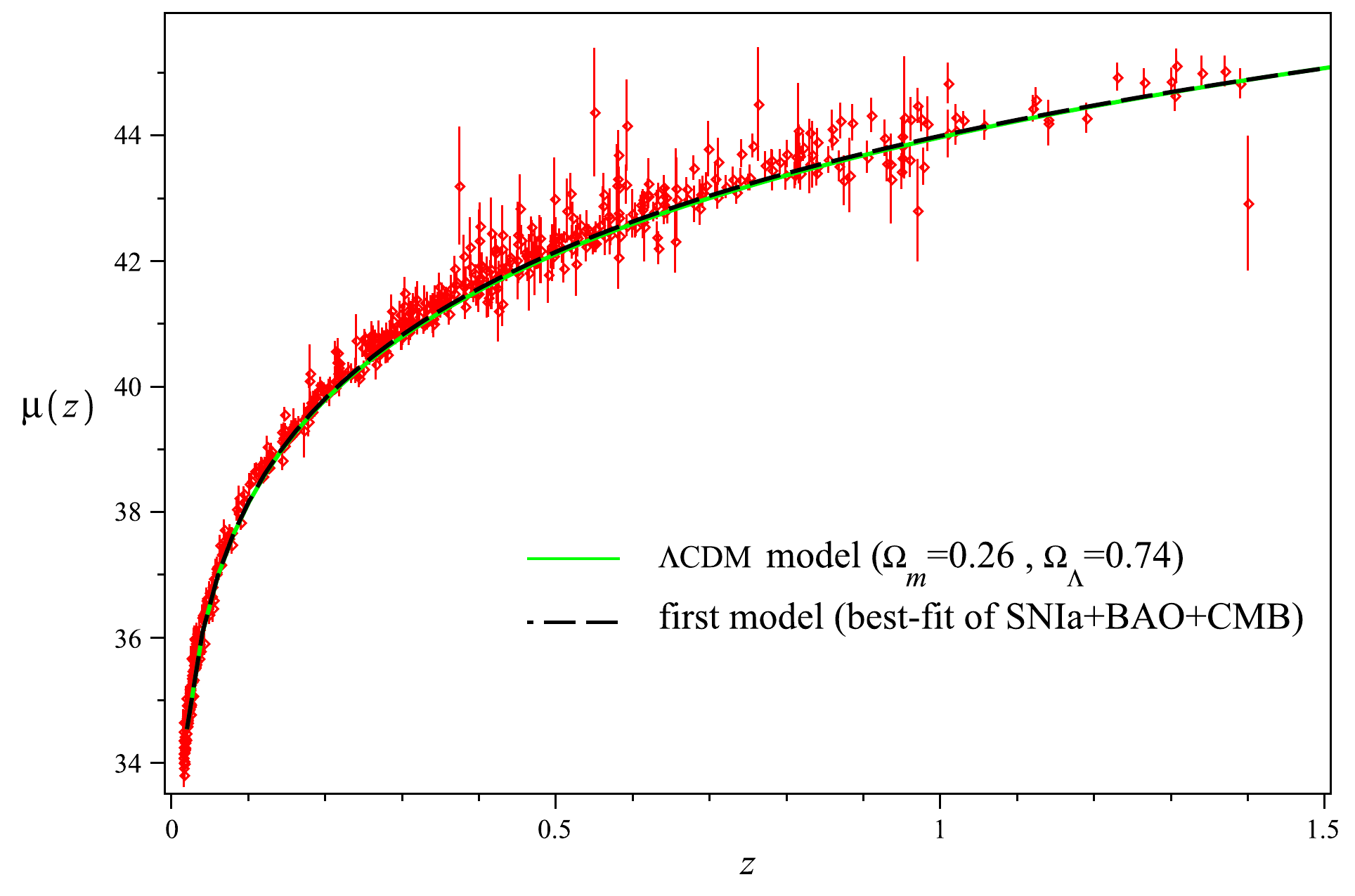}
\includegraphics[width=0.48\textwidth]{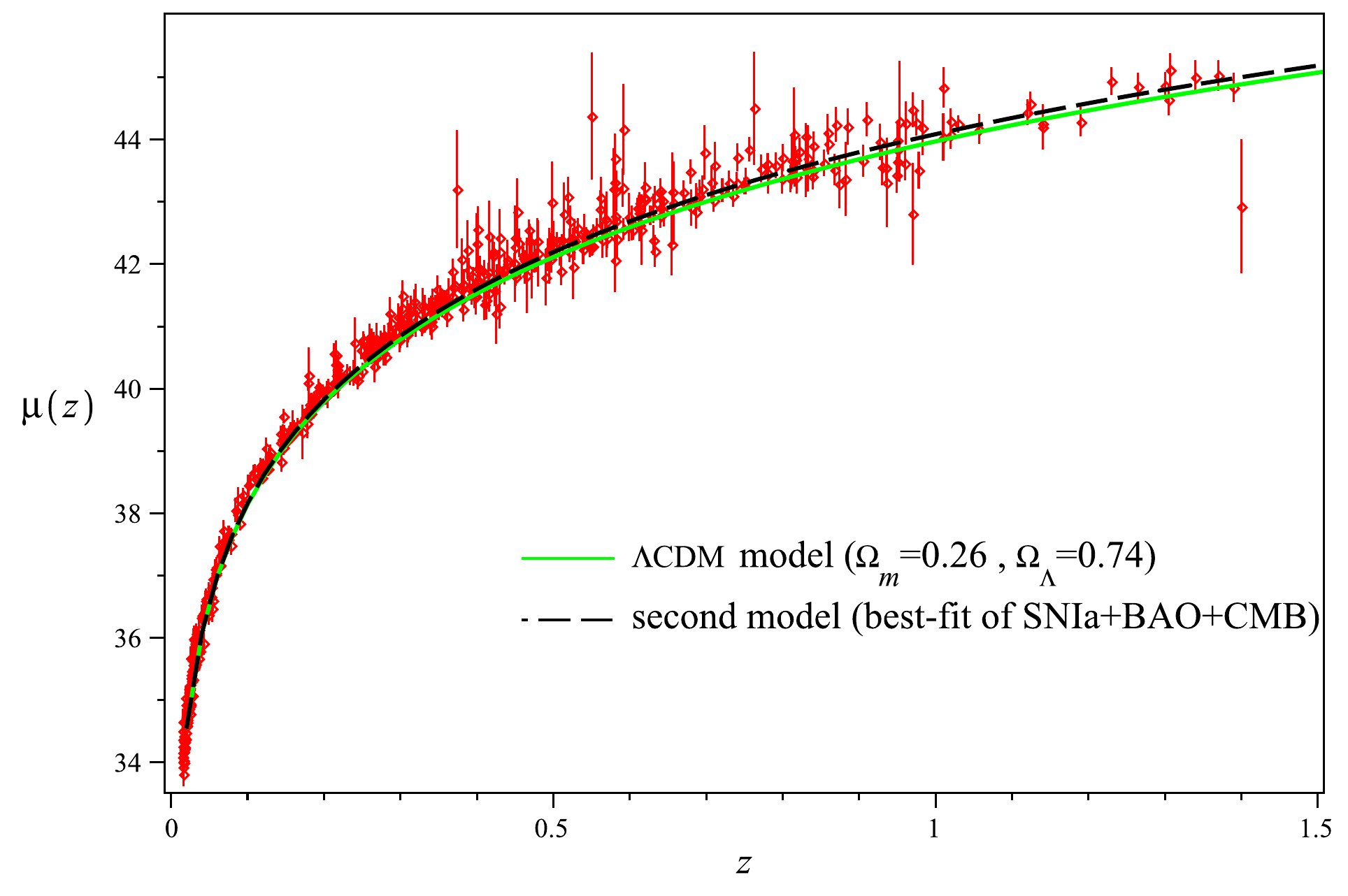}
\caption{The distance modulus, $\mu(z)$, with respect to $z$ expected for the standard concordance model
$\Lambda$CDM with $\Omega_\Lambda=0.74$ and $\Omega_{m0}=0.26$ (solid line), and for our models (dashed line) with the values of the best-fit coming
from SnIa+BAO+CMB. Top panel: the first model. Bottom panel: the second model.}\label{fig7}
\end{figure}

The distance modulus, $\mu(z)$, plotted in FIG. (\ref{fig7}), for both models with the best fitted parameter values are compared to the $\Lambda$CDM model by combining Sne Ia, BAO and CMB observational data.

\subsection{Dark torsion EoS parameter}

\begin{figure}
\centering
\includegraphics[width=0.38\textwidth]{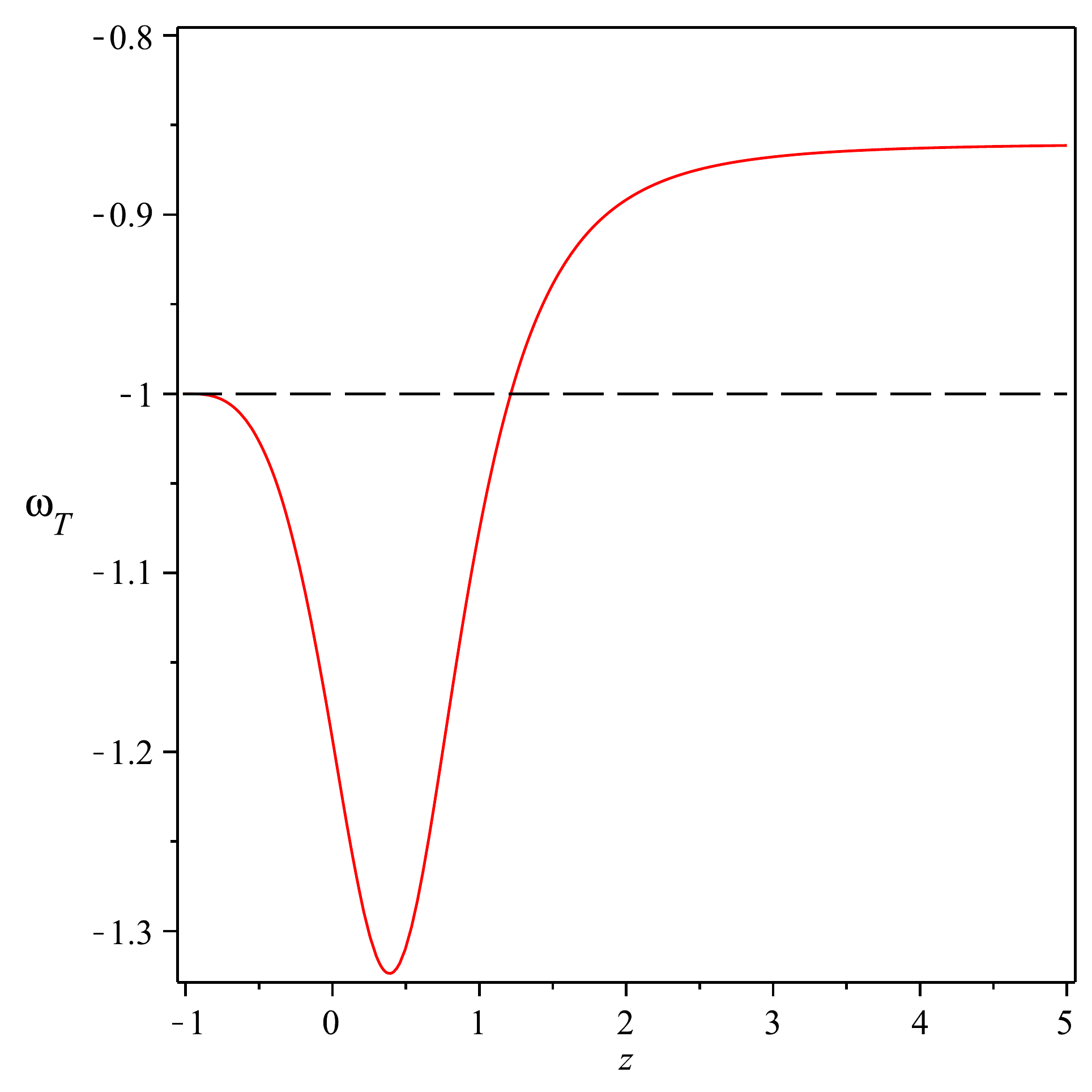}
\includegraphics[width=0.38\textwidth]{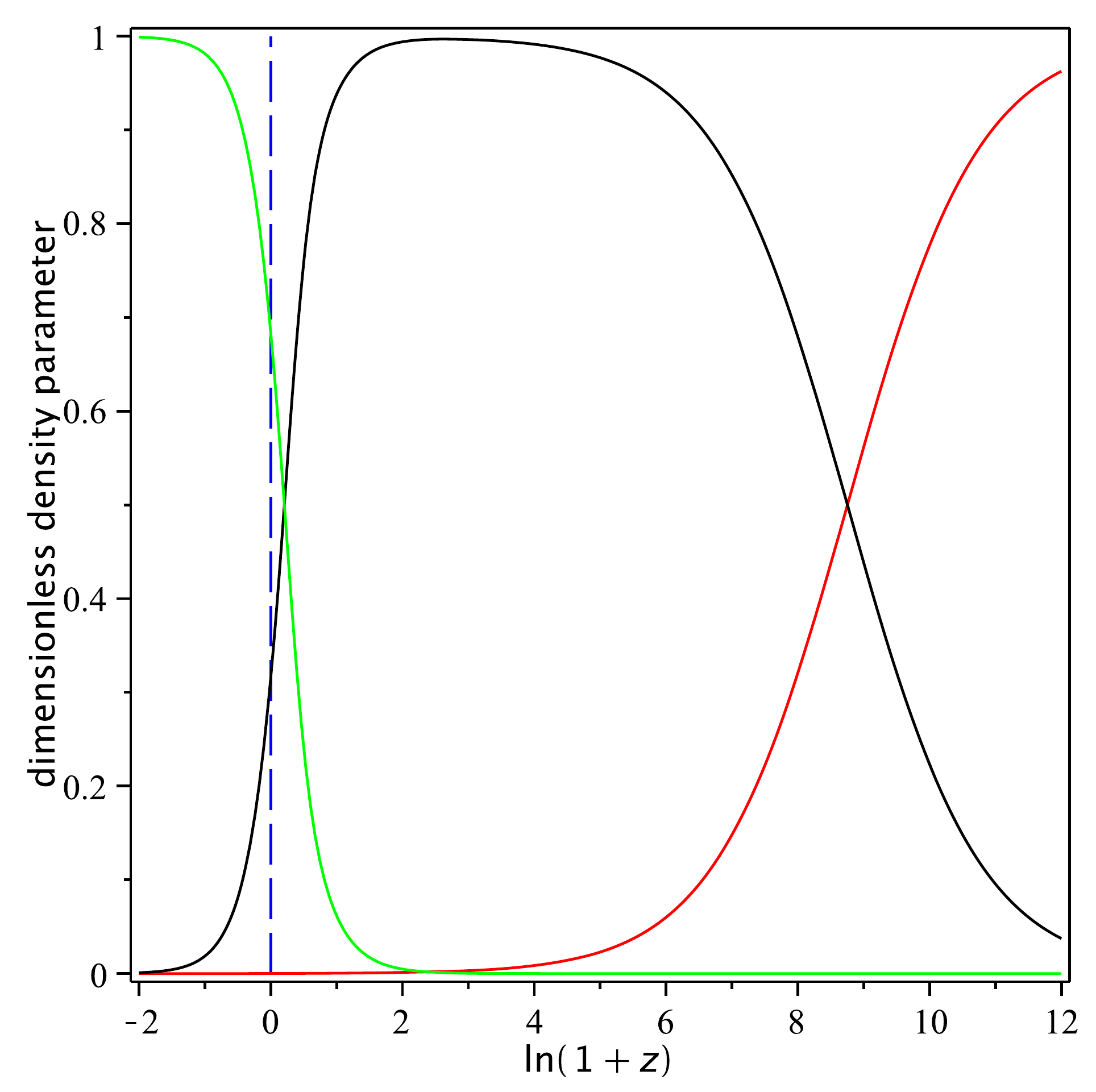}
\caption{The trajectories of the dark torsion EoS parameter (top panel) and the trajectories of the dimensionless density parameters for the DT (green),
radiation (red) and matter (black) (bottom panel) with the best fitted values of $n$ and $\Omega_{m0}$ in the fist model. }\label{fig1}
\end{figure}

Using the best-fitted constrained model parameters in the last section, we can now discuss the behavior of dark torsion EoS parameters in the models.
In the first model, FIG. (\ref{fig1}), bottom), shows the dynamics of the dimensionless density parameters for radiation, matter and DT in the universe.
It reveals that the universe in this model has a long enough period of radiation domination to give the correct primordial nucleosynthesis and
radiation-matter equality and a matter dominated phase. In other words, the usual early universe behavior can
be successfully obtained to agree with the primordial nucleosynthesis and the cosmic microwave background constraints.
It also shows that for this model the DT start dominating the universe at about $z\simeq 0.23$. \\

In addition, in FIG. (\ref{fig1}) top), the evolutionary curves of the dark torsion EoS parameter for the best fitted values of $n$ and $\Omega_{m0}$ are shown. One can see that the dark torsion EoS parameter crosses the phantom divide line from the values greater than $-1$ (non-phantom phase) to smaller than $-1$ (phantom phase) in the past and become tangent to $-1$ in the future. The result is in consistent with the one obtained
in \cite{Wu3} where the universe in the $f(T)$ theory finally enters a de Sitter expansion phase. Also, it can be easily shown that the crossing occurs at the red shift $z$ where Eq.(\ref{omegacross11}) is satisfied and the constraint (\ref{omegacross2}) is related to entering a de Sitter expansion phase.

\begin{figure}
\centering
\includegraphics[width=0.38\textwidth]{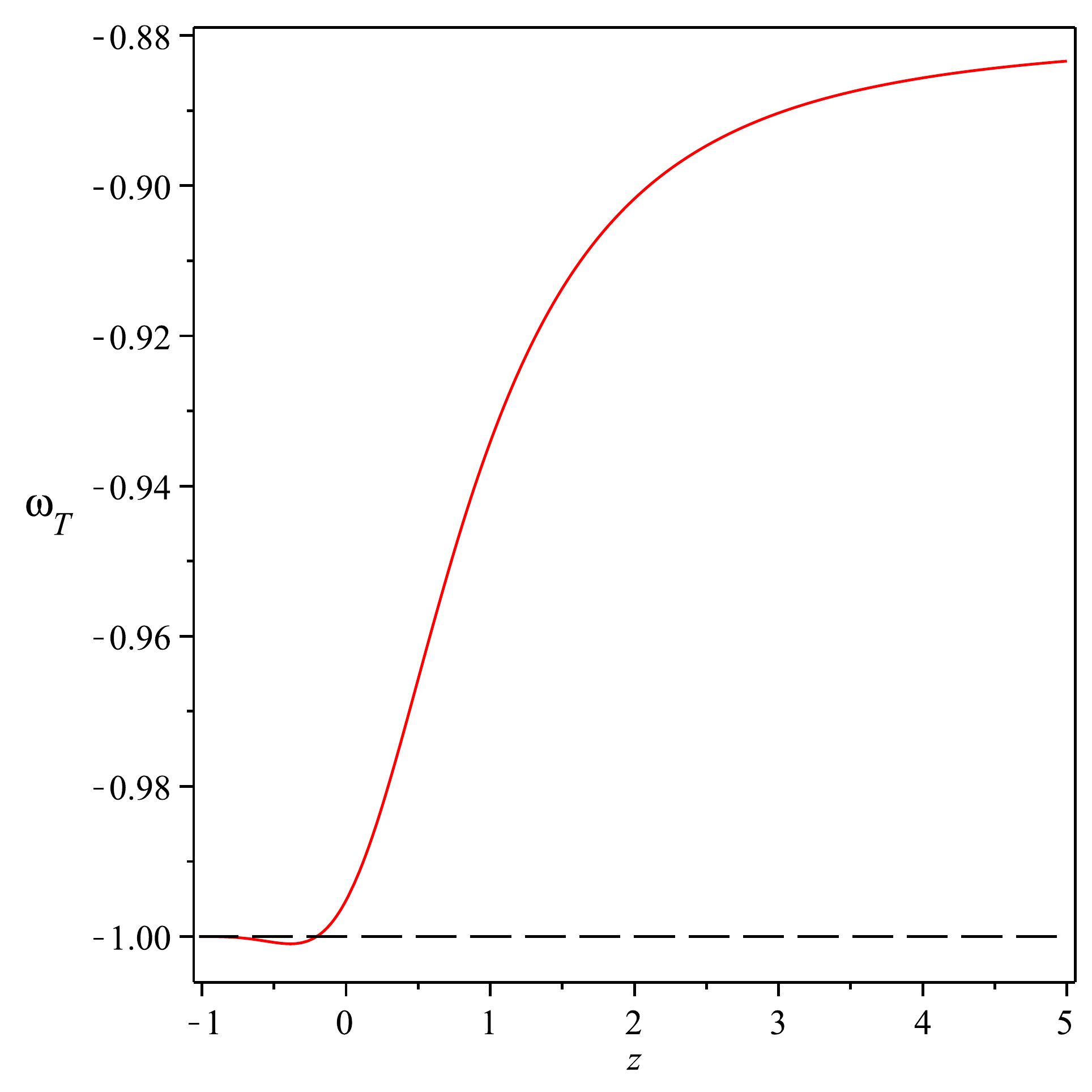}
\includegraphics[width=0.38\textwidth]{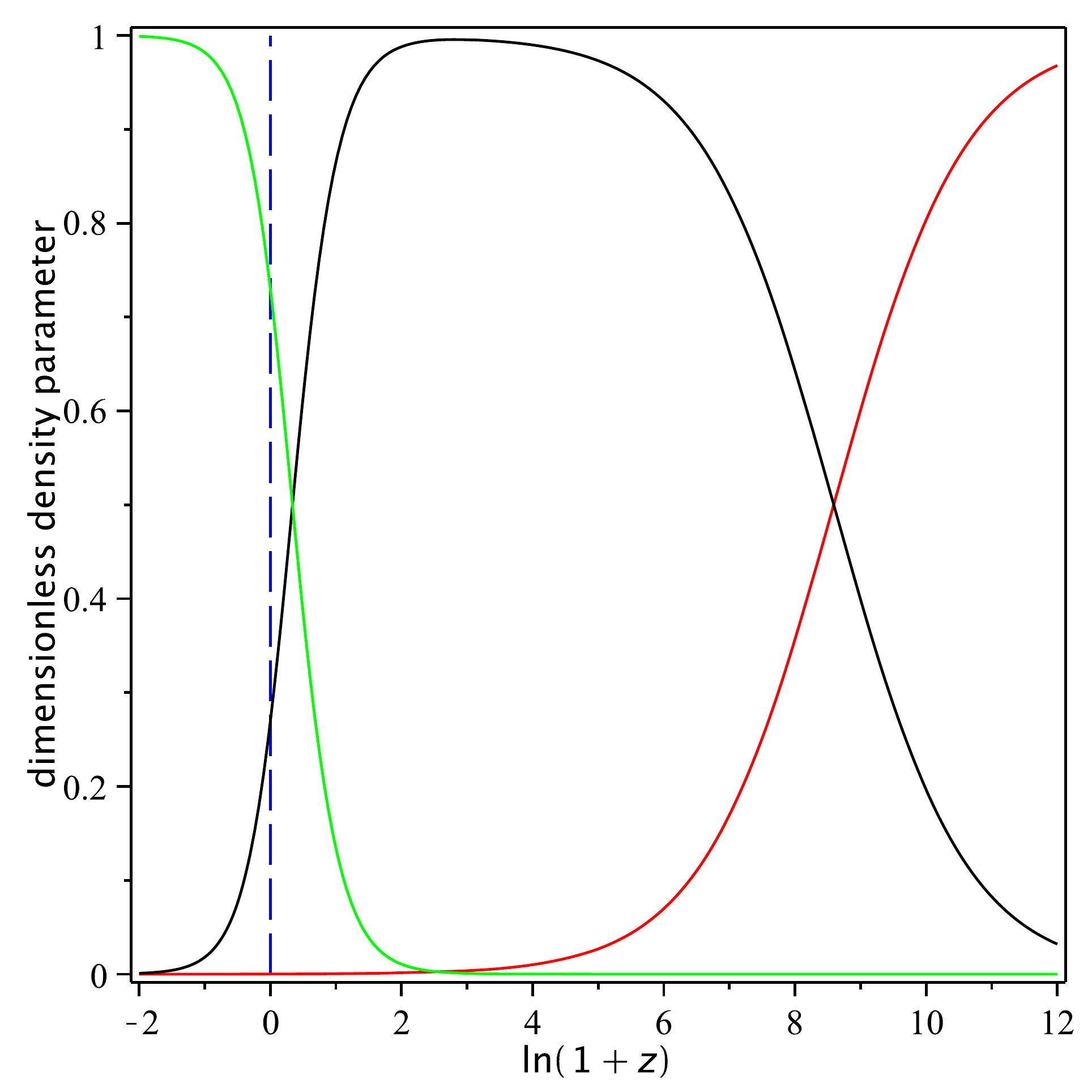}
\caption{The trajectories of the dark torsion EoS parameter (top panel) and the trajectories of the dimensionless density parameters for the DT (green),
radiation (red) and matter (black) (bottom panel) with the best fitted values of $n$, $\Omega_{m0}$ and $p$ in the second model.}\label{fig3}
\end{figure}

For the second model, in FIG. (\ref{fig3}) bottom)  we show the trajectories of the dimensionless density parameters and the dark torsion EoS parameter
for the best fitted model parameters of $p$,  $n$, and $\Omega_{m0}$ and the DT start dominating the universe at about $z\simeq 0.4$.
One can see in the top panel that the dark torsion EoS parameter crosses the phantom divide line from the values greater than $-1$ to smaller than $-1$ in the future and finally become tangent to $-1$ in the near future. Also, in this figure one can check that the conditions (\ref{omegacross33}) or (\ref{omegacross4}) satisfies when the crossing occurs in the early future and the model enters a de Sitter expansion phase in the late future.

\subsection{Total EoS parameter}

In the pervious section we examined the dynamics of the dark torsion EoS parameter which in the first model shows phantom crossing in the past plus de Sitter phase in the future and in the second model shows phantom crossing in the early future and de Sitter phase in the late future. However, the universe is filled with the CDM and radiation and a physical model requires the inclusion of correction terms in the formulation.  One can rewrite the Friedmann equations (\ref{friedmann}) and (\ref{acceleration}) as
\begin{eqnarray}
  T &=& -16\pi G\rho_{tot}\label{friedtotal1} \\
  2\frac{\dot{T}}{\sqrt{-6T}} +T &=& 16\pi Gp_{tot}\label{friedtotal2}
\end{eqnarray}
where $\rho_{tot} = \rho + \rho_{T}$ and $p_{tot} = p + p_{T}$ and the energy density, $\rho$, is for CDM and radiation filled the universe. Simply, the total EoS parameter in terms of the redshift $z$ as
\begin{equation}\label{omegatot}
    \omega_{tot}=-1+\frac{(1+z)}{3T}\frac{dT}{dz}\cdot
\end{equation}
Using the best fitted model parameters obtained from $\chi^2$ method, one observes the evolution of the total EoS parameter $\omega_{tot}$ as a function of $z$ for the above two models.

\begin{figure}
\centering
\includegraphics[width=0.48\textwidth]{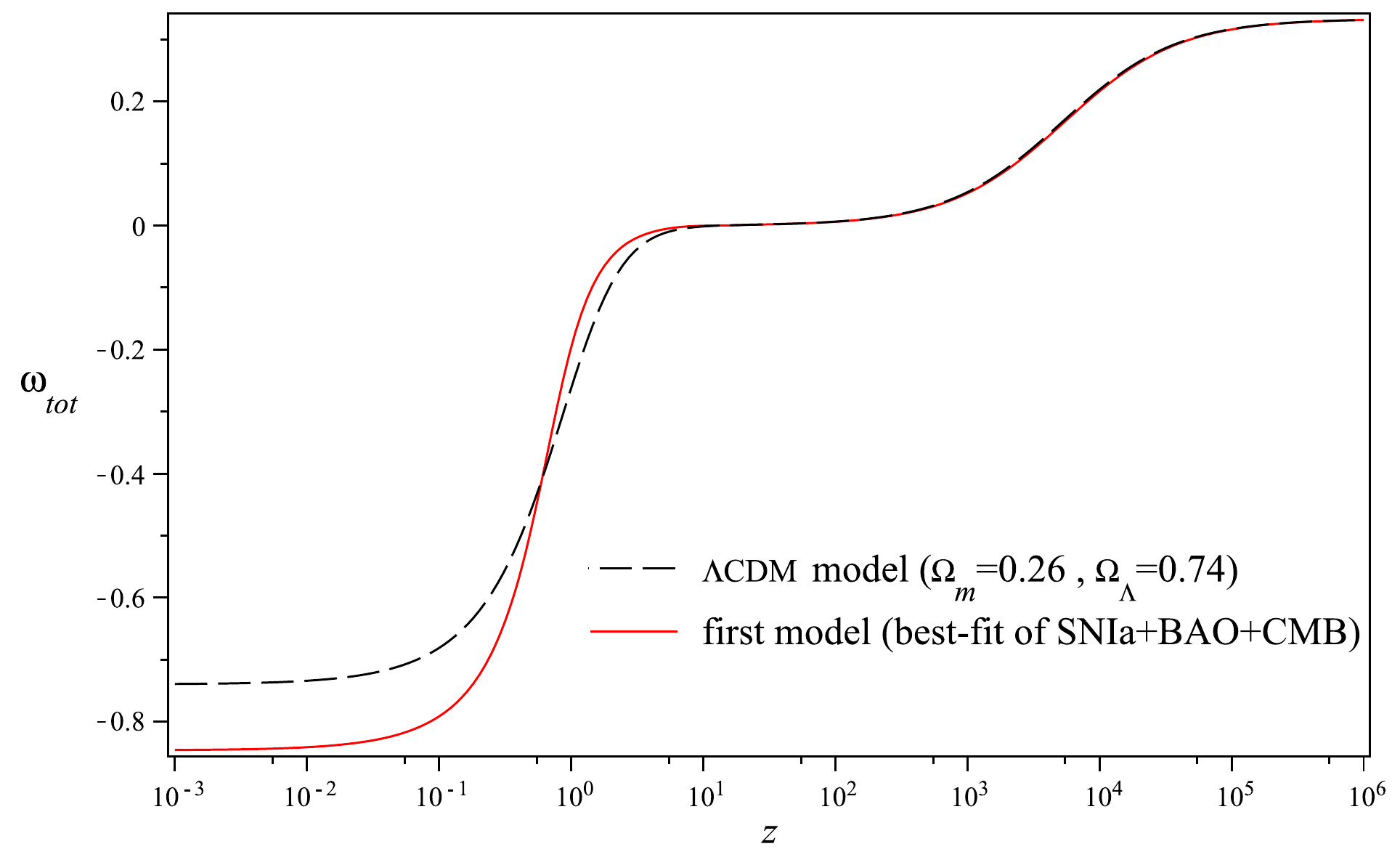}
\includegraphics[width=0.48\textwidth]{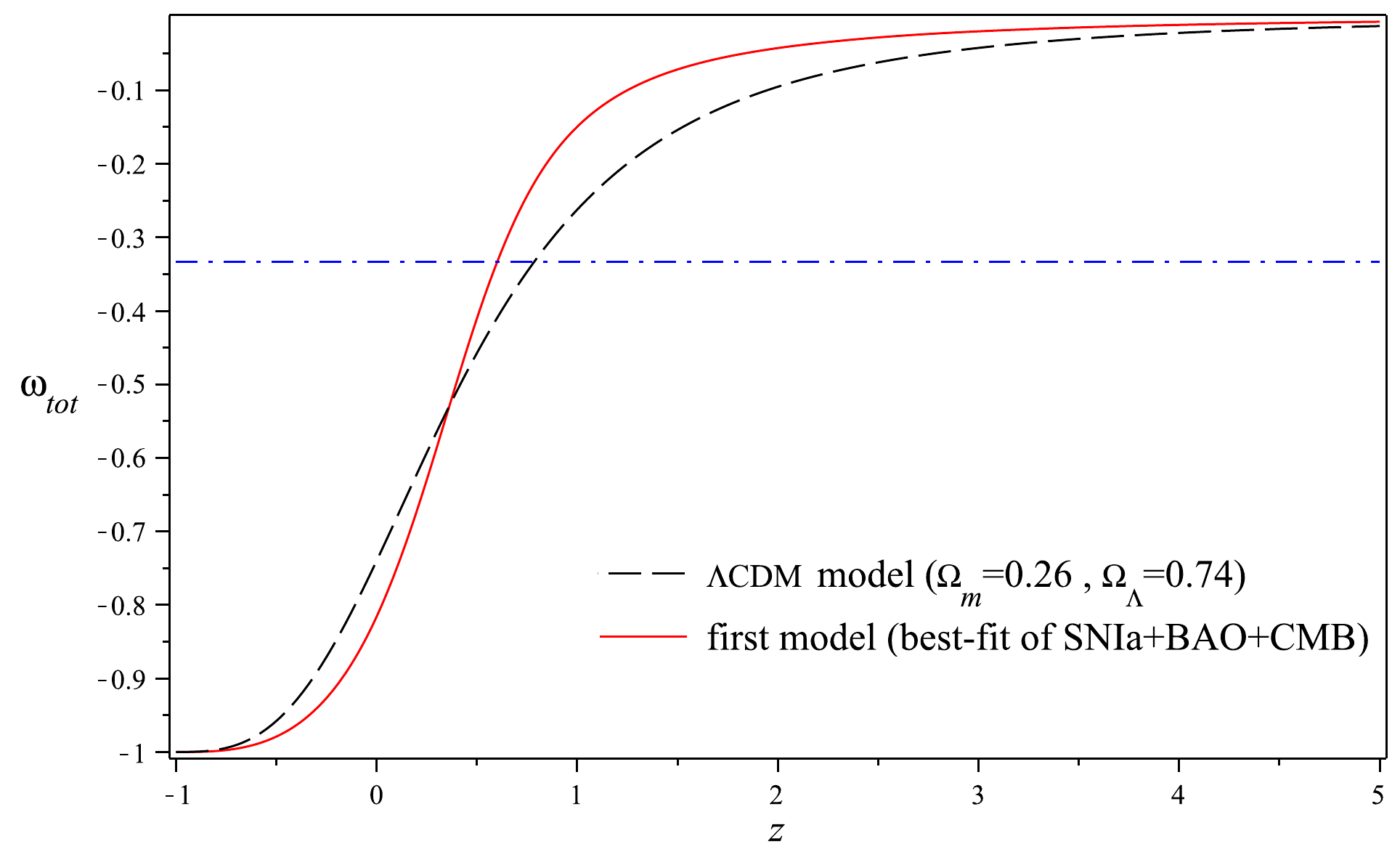}
\caption{The trajectories correspond to the total EoS parameter as a function of $z$ expected for the standard concordance model
$\Lambda$CDM with $\Omega_\Lambda=0.74$ and $\Omega_{m0}=0.26$ (dashed line), and for our models (solid lines) with the values of the best-fit coming
from SNIa+BAO+CMB, for the first model in two different ranges}\label{fig8}
\end{figure}

FIG. (\ref{fig8}) top) shows the last three phases of the evolution of the universe, i.e., the radiation dominated ($\omega_{tot}=1/3$),
the matter dominated ($\omega_{tot}=0$) and the late time acceleration for the first model in comparison with the $\Lambda$CDM model.
In more details to observe the behavior of the $\omega_{tot}$ in the near past and future, the FIG. (\ref{fig8}) bottom) shows that the universe transits from deceleration era to acceleration era when $\omega_{tot}=-1/3$ at about $z\sim0.61$ and approaches a de Sitter phase in the future where $\omega_{tot}\rightarrow -1$. We also see that the current value of $\omega_{tot}$ at $z=0$ is about $\omega_{tot}\simeq -0.82$ whereas for $\Lambda$CDM model is about $\simeq -0.74$.\\

\begin{figure}
\centering
\includegraphics[width=0.48\textwidth]{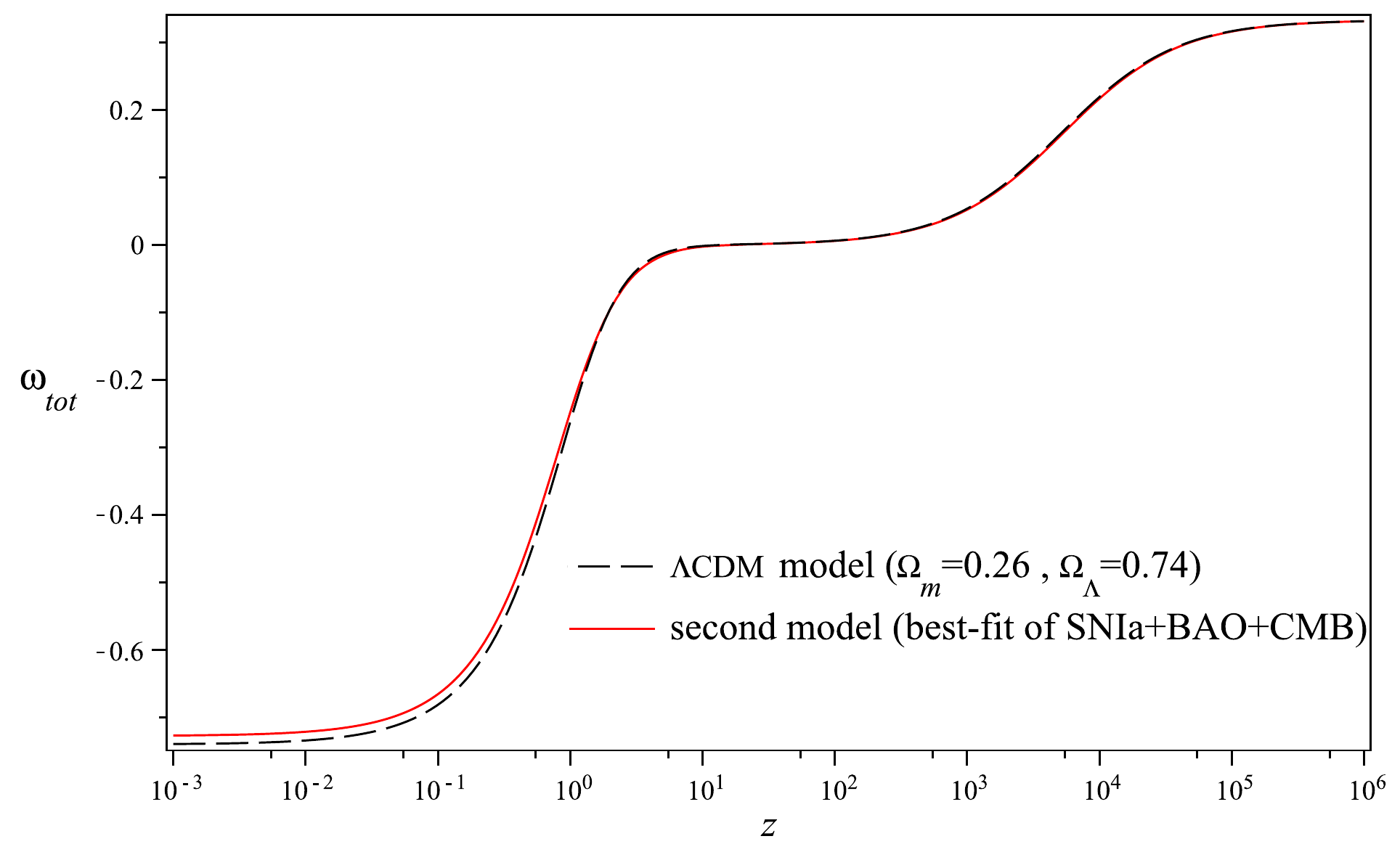}
\includegraphics[width=0.48\textwidth]{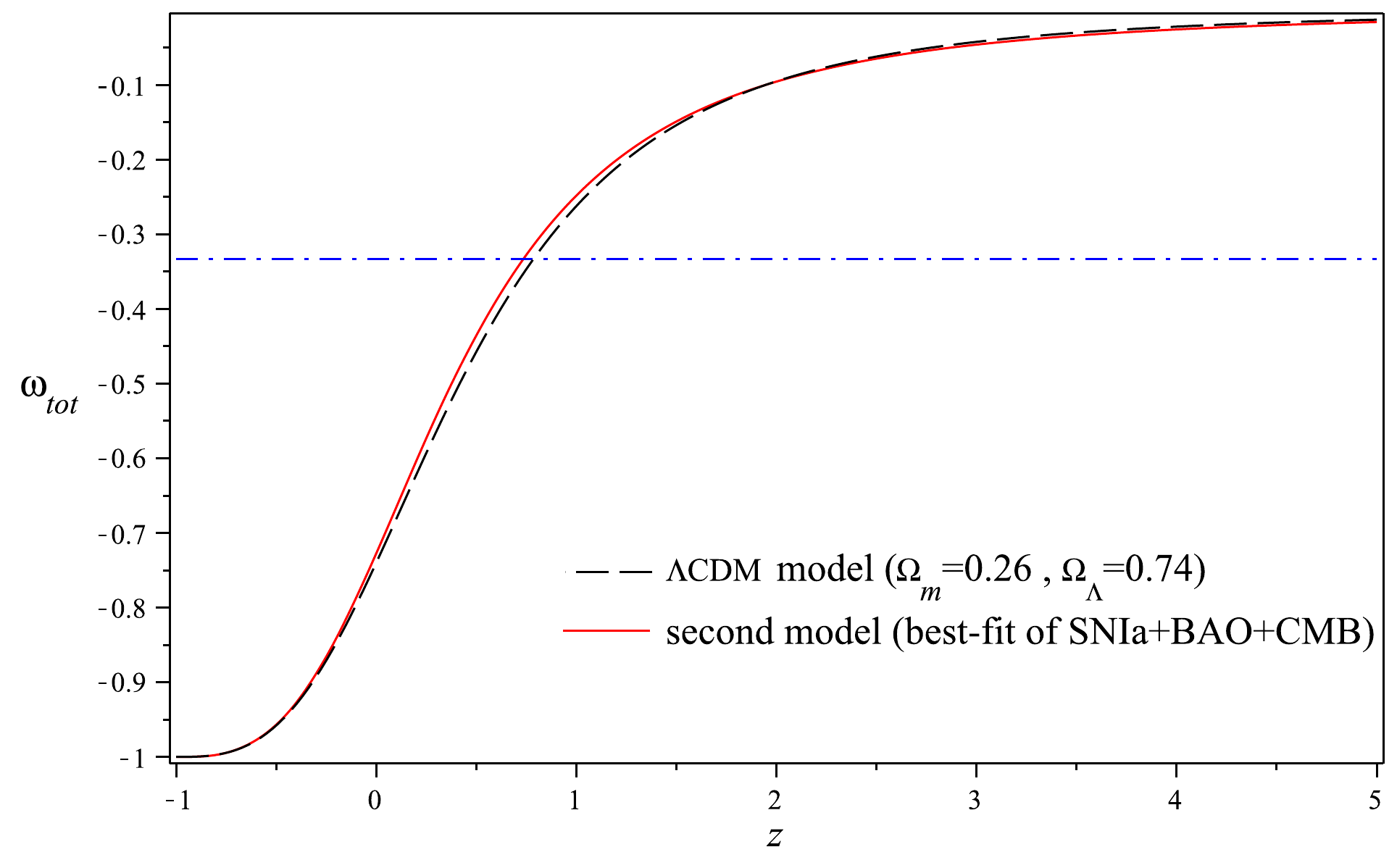}
\caption{The trajectories correspond to the total EoS parameter as a function of $z$ expected for the standard concordance model
$\Lambda$CDM with $\Omega_\Lambda=0.74$ and $\Omega_{m0}=0.26$ (dashed line), and for our models (solid lines) with the values of the best-fit coming
from SNIa+BAO+CMB, for the second model in two different ranges.}\label{fig19}
\end{figure}

Also, FIG. (\ref{fig19}) top) shows the last three phases of the evolution of the universe, i.e., the radiation dominated ($\omega_{tot}=1/3$),
the matter dominated ($\omega_{tot}=0$) and the late time acceleration for the second model in comparison with the $\Lambda$CDM model.
For the behavior of the $\omega_{tot}$ in the near past and future, the FIG. (\ref{fig19}) bottom) shows that the universe transits from deceleration era to acceleration era when $\omega_{tot}=-1/3$ at about $z\sim 0.74$ and approaches a de Sitter phase in the future where $\omega_{tot}\rightarrow -1$. The current value of $\omega_{tot}$ in this model is about $\omega_{tot}\simeq -0.73$ whereas for $\Lambda$CDM model it is about $\simeq -0.74$.\\

\section{Summary and Remarks}

In this manuscript, the theory of $T +f(T )$ based on TEGR where torsion
is the geometric object describing gravity instead of
the curvature and the 2nd order differential equations that is remarkably simpler than $f(R)$ theories is investigated for two different
$f(T)$ models. Our purpose in this work is, by using
numerical methods, to reproduce
the recently detected acceleration of the universe and the dynamics of the EoS parameter for DT
without resorting to the DE. Our work is distinguished from the work given in \cite{Wu} by an analytic discussion on model equations for phantom crossing behavior for the given $f(T)$ models. In addition the parameters in the model are best fitted with the observational data in advance, before numerical computation of the EoS parameters for the models. This gives us a more accurate picture of the universe and physically meaningful solution for the derived parameters.

Our original motivation for studying $f(T)$ gravity and considering the specific models as an alternative to conventional dark energy in general relativistic cosmology is to interpret the current universe acceleration and reproduce a dynamical dark energy in agreement with the observations. Our $f(T)$ models satisfy the conditions at the high redshift, consistent with the primordial nucleosynthesis and cosmic microwave background constraints. Furthermore, the usual general relativity can be retrieved at low energy limit.

We tested the observational viability of the models by
using the most recent SNe Ia data combined
with the information coming from BAO peak and CMB
shift parameter in order to find constraints on the model parameters, $n$, $p$ and $\Omega_{m0}$.
The numerical analysis indicates that these two
types of $f(T)$ theories are compatible with observations. At $ 68.3\% $ confidence for the first model we found that the parameter's values lie in the
ranges  $n=[0.9, 1.3]$ and $\Omega_{m0}=[0.288, 0.345]$. The model with the best fitted values minimizing the $\chi^2$  that combines
SNe Ia+BAO+CMB data given by $n = 1.14$ and $\Omega_{m0}=0.316$. Furthermore, for the second model, at $ 68.3\% $ confidence we found that the parameter's values lie in the ranges  $n=[0.79, 1.44]$,
$\Omega_{m0}=[0.243, 0.302]$ and $p=[-0.48, 0.17]$. The model
with the best-fit values minimizing the $\chi^2$  that combines
SNe Ia+BAO+CMB data for this model given by $n = 1.12$, $p=0.05$ and
$\Omega_{m0}=0.269$.

With the best fitted model parameters for the distance modulus and in comparison with the observational data we calculate the dark torsion EoS parameter and also dimensionless density parameters for the two models. In the first model, the phantom crossing occurs in the past, while currently in phantom era and approaches a de Sitter phase in future when $z=-1$ and universe become infinite. For the second model the universe is currently in quintessence era, undergoes phantom crossing in near future and again approaches a de Sitter phase later in future when $z=-1$ and universe become infinite.

However, in reality, in addition to the repulsive DT, the universe is also filled with CDM and radiation. Thus the dynamics of the universe has to be explained with the total EoS parameter, $\omega_{tot}$. The computation given for $\omega_{tot}$ exhibits the last three phases of cosmological evolution:
radiation era, matter era and late acceleration where in the last
stage transition from deceleration to acceleration occurs for the first model at $z\sim 0.61$, the second model at $z\sim 0.74$ and for $\Lambda$CDM model at $z \sim 0.75$. Also, from the graphs the current values for $\omega_{tot}$ for the first model is $\omega_{tot} \sim -0.82$ and the second model is $\omega_{tot} \sim -0.73$ whereas for the $\Lambda$CDM model is $\sim -0.74$. One can conclude from findings that the second model is more compatible with the $\Lambda$CDM model \cite{Mel}. We also find that the values obtained for the total EoS parameters in both models are within the $1 \sigma$ confidence region of current EoS parameter in the WMAP 5-year results \cite{Kom2}.

\nocite{*}
\bibliographystyle{spr-mp-nameyear-cnd}
\bibliography{biblio-u1}

\begin{thebibliography}{}

\bibitem[(Riess 1998)]{Riess} Riess, A. G., et al. 1998, J. Astron. 116, 1009
\bibitem[(Perlmutter 1999)]{Perlm} Perlmutter, S., et al. 1999, Astrophys. J. 517, 565
\bibitem[(Spergel 2003)]{Sper} Spergel, D. N., et al. 2003, ApJS, 148, 175
\bibitem[(Spergel 2007)]{Sper2} Spergel, D. N., et al. 2007, ApJS, 170, 377S
\bibitem[(Tegmark 2004)]{Teg} Tegmark, M., et al. 2004, Phys. Rev. D. 69, 103501
\bibitem[(Eisenstein 2005)]{Eis} Eisenstein, D. J., et al. 2005, Astrophys. J. 633, 560
\bibitem[(Carroll 2001)]{Car} Carroll, S. M., 2001, Living Rev. Rel. 4, 1
\bibitem[(Copeland 2006)]{Cop} Copeland, E. J., Sami M. \&  Tsujikawa, S. 2006, Int. J. Mod. Phys. D. 15, 1753
\bibitem[(Yang 2010)]{Yang1} Yang, R. J. \& Zhang S. N. 2010, Mon. Not. R. Astron. Soc. 407, 1835
\bibitem[(Feng 2005)]{Feng} Feng, B., Wang X. L. \&  Zhang, X. M. 2005, Phys. Lett. B. 607, 35
\bibitem[(Caldwell 1998)]{Cald} Caldwell, R. R., Dave R. \&  Steinhardt, R. J. 1998, Phys. Rev. Let. 80, 1582

\bibitem[(Armendariz 2001)]{Arm} Armendariz-Picon, C., Mukhanov V. \&  Steinhardt, P. J. 2001, Phys. Rev. D 63, 103510

\bibitem[(Padmanabhan 2002)]{Pad} Padmanabhan, T., 2002, Phys. Rev. D 66, 021301



\bibitem[(Sen 2005)]{Sen} Sen, A., 2005, Phys. Scripta. T. 117, 70

\bibitem[(Caldwell 2002)]{Cald2} Caldwell, R. R., 2002, Phys. Lett. B. 545, 23

\bibitem[(Elizadle 2004)]{Eli} Elizadle, E., Nojiri, S. \&  Odintsov, S. D. 2004, Phys. Rev. D 70, 043539




\bibitem[(Kamenshchik 2001)]{Kam} Kamenshchik, A., Moschella, U. \&  Pasquier, V. 2001, Phys. Lett. B. 511, 265



\bibitem[(Bengochea 2011)]{Ben} Bengochea, G. R., 2011, Phys. Lett. B. 695, 405

\bibitem[(Cohen 1999)]{Cohen} Cohen, A. G., Kaplan, D. B. \&  Nelson, A. E. 1999, Phys. Rev. Let. 82, 4971

\bibitem[(Li 2011)]{Li} Li, B., Sotiriou, T. P. \&  Barrow, J. D. 2011, Phys. Rev. D 83, 064035

\bibitem[(Sotiriou 2011)]{Sotiriou} T. P. Sotiriou, B. Li, J. D. Barrow, 2011 Phys. Rev. D 83:104030

\bibitem[(Wei 2008)]{Wei} Wei, H. \&  Cai, R. G. 2008, Phys. Lett. B. 663, 1
\bibitem[(Wei$^{a}$ 2008)]{Wei2} Wei$^{a}$, H. \&  Cai, R. G. 2008, Phys. Lett. B. 660, 113
\bibitem[(Gao 2009)]{Gao} Gao, C., Wu, F., Chen, X. \&  Shen, Y. G. 2009, Phys. Rev. D 79, 043511


\bibitem[(Elizalde 2011)]{Eliz1}E. Elizalde, S. D. Odintsov, L. Sebastiani, S. Zerbini, arXiv:1108.6184 [gr-qc]

\bibitem[(Elizalde$^{a}$ 2011)]{Eliz2}E. Elizalde, S. Nojiri, S.D. Odintsov, L. Sebastiani, S. Zerbini, Phys.Rev.D83:086006,2011, arXiv:1012.2280 [hep-th];

\bibitem[(Cognola 2008)]{Cogn}G. Cognola, E. Elizalde, S. Nojiri, S.D. Odintsov, L. Sebastiani, S. Zerbini, Phys.Rev.D77:046009,2008, arXiv:0712.4017 [hep-th]





\bibitem[(Sotiriou 2010)]{Sot} Sotiriou, T. P. \&  Faraoni, V. 2010, Rev. Mod. Phys. 82, 451
\bibitem[(Chiba 2003)]{Chiba} Chiba, T., 2003, Phys. Lett. B. 575, 1
\bibitem[(Olmo 2005)]{Olmo} Olmo, G., 2005, Phys. Rev. Let. 95, 261102

\bibitem[(Amendola 2007)]{Amen} Amendola, L., et al. 2007, Phys. Rev. Let. 98, 131302
\bibitem[(Amendola$^{a}$ 2007)]{Amen2} Amendola$^{a}$, L., et al. 2007, Phys. Rev. D 75, 083504
\bibitem[(Amarzguioui 2006)]{Amar} Amarzguioui, M., et al. 2006, Astron. and Astrophys. 454, 707
\bibitem[(Fay 2007)]{Fay} Fay, S., Tavakol, R. \&  Tsujikawa, S. 2007, Phys. Rev. D 75, 063509
\bibitem[(Santos 2008)]{San} Santos, J., et al. 2008, Phys. Lett. B. 669, 14


\bibitem[(Nojiri 2011)]{Nojiri1}S. Nojiri, S. D. Odintsov, 	Phys.Rept.505:59-144,2011, arXiv:1011.0544

\bibitem[(Nojiri 2005)]{Nojiri2}S. Nojiri, S. D. Odintsov, 	Phys.Lett.B631:1-6,2005, hep-th/0508049

\bibitem[(Nojiri 2006)]{Nojiri3}S. Nojiri, S.D. Odintsov, 	ECONF C0602061:06,2006; Int.J.Geom.Meth.Mod.Phys.4:115-146,2007, hep-th/0601213




\bibitem[(Felice 2010)]{Felice} Felice, A. D., Mota, D. F. \&  Tsujikawa, S. 2010, Phys. Rev. D 81, 023532




\bibitem[(Nojiri 2004)]{Nojiri4}S. Nojiri, S. D. Odintsov, Phys.Lett.B599:137-142,2004, astro-ph/0403622.



\bibitem[(Farajollahi 2010)]{Faraj} Farajollahi, H., Farhoudi, M. \&  Shojaie, H. 2010, Int. J. Theor. Phy. 49, 10, 2558
\bibitem[(Zuntz 2010)]{Zuntz} Zuntz, J., Zlosnik, T. G., Bourliot, F., Ferreira, P. G. \&  Starkman, G. D. 2010, Phys. Rev. D 81, 104015
\bibitem[(Camera 2010)]{Camera} Camera, M. L., 2010, Mod. Phys. Lett. A. 25, 781-792
\bibitem[(Nojiri 2003)]{Noj} Nojiri, S. \&  Odintsov, S. D. 2003, Phys. Rev. D 68, 123512
\bibitem[(Nojiri 2006)]{Noj2} Nojiri, S. \&  Odintsov, S. D. 2006, Phys. Rev. D 74, 086005
\bibitem[(Nojiri 2007)]{Noj3} Nojiri, S. \&  Odintsov, S. D. 2007, Int. J. Geom. Meth. Mod. Phys. 4, 115-146
\bibitem[(Abdalla 2005)]{Abd} Abdalla, M. C. B., Nojiri, S. \&  Odintsov, S. D. 2005, Class. Quant. Grav. 22, L35
\bibitem[(Nojiri 2008)]{Noj4} Nojiri, S. \&  Odintsov, S. D. 2008, Phys. Rev. D 77, 026007
\bibitem[(Einstein 1928)]{Einstein} Einstein, A., 1928, Sitz. Preuss. Akad. Wiss. p. 217; ibid p. 224
\bibitem[(Einstein 2005)]{Einstein2} Einstein, A., 2005, translations of Einstein papers by Unzicker, A. \& Case, T., (arXiv:physics/0503046).
\bibitem[(Hoff da Silva 2010)]{Hoff} Hoff da Silva, J. M. \&  da Rocha, R. 2010, Phys. Rev. D 81, 024021
\bibitem[(Hayashi 1979)]{Hay} Hayashi, K. \&  Shirafuji, T. 1979, Phys. Rev. D 19, 3524
\bibitem[(Ferraro 2008)]{Fer} Ferraro, R. \&  Fiorini, F. 2008, Phys. Rev. D 78, 124019
\bibitem[(Ulhoa 2010)]{Ulhoa} Ulhoa, S. C., da Rocha Neto, J. F. \&  Maluf, J. W. 2010, Int. J. Mod. Phys. D. Vol. 19, No. 12, 1925-1935
\bibitem[(Nashed 2010)]{Nash} Nashed, G. G. L., 2010, Int. J. Mod. Phys. A. Vol. 25, No. 14, 2883-2895
\bibitem[(Sharif 2010)]{Sharif} Sharif, M. \&  Taj, S. 2010, Mod. Phys. Lett. A. 25, 221-232
\bibitem[(Lucas 2009)]{Lucas} Lucas, T. G., Obukhov, Y. N. \&  Pereira, J. G. 2009, Phys. Rev. D 80, 064043
\bibitem[(Ferraro 2007)]{Fer2} Ferraro, R. \&  Fiorini, F. 2007, Phys. Rev. D 75, 084031
\bibitem[(Poplawski 2010)]{Pop} Poplawski, N. J., 2010, Phys. Lett. B. 694, 181-185
\bibitem[(Wu 2011)]{Wu} Wu, P. \&  Yu, H. 2011, Eur. Phys. J. C. 71, 1552
\bibitem[(Wu 2010)]{Wu2} Wu, P. \&  Yu, H. 2010, Phys. Lett. B. 693, 415-420
\bibitem[(Ao 2010)]{Ao} Ao, X. C., Li, X. Z. \&  Xi, P. 2010, Phys. Lett. B. 694, 186-190

\bibitem[(Bengochea 2009)]{Ben2} Bengochea, G. \&  Ferraro, R. 2009, Phys. Rev. D 79, 124019
\bibitem[(Wu$^{a}$ 2010)]{Wu3} Wu$^{a}$, P. \&  Yu, H. 2010, Phys. Lett. B. 692, 176-179

\bibitem[(Yang 2010)]{Yang} Yang, R. J., 2010, (arXiv:1007.3571v2)





\bibitem[(Bamba 2010)]{Bamba} Yang, R. J., 2011 J. Cosm. and Astro. Phys. 1101:021




\bibitem[(Shie 2008)]{Shie}K. F. Shie, J. M. Nester, and H. J. Yo, 2008, Phys. Rev. D 78:023522

\bibitem[(Poplawski 2010)]{Poplawski}N. J. Poplawski, 2010 plb, Vol. 694, No. 3, pp. 181–185



\bibitem[(Cai 2011)]{Cai}Y. F. Cai, et al., 2011, Class. Quantum Grav. 28 215011

\bibitem[(Daouda 2011)]{Daouda}M. H. Daouda, M. E. Rodrigues and M. J. S. Houndjo, arXiv:1108.2920v3

\bibitem[(Bamba 2010)]{Bamba1}K. Bamba and C.Q. Geng, 	arXiv:1109.1694v1

\bibitem[(Huang 2008)]{Huang}C. G. Huang, H. Q Zhang and H. Y. Guo, 2008 J. Cosm. and Astro. Phys. 10 010

\bibitem[(Li 2009)]{Li1}X. Z. Li, C. b. Sun and P. Xi, 2009 J. Cosm. and Astro. Phys. 04 015

\bibitem[(Esposoto 1991)]{Esposoto}G. Esposoto,1991 NuovoCim.B104:199-212,1989; Erratum-ibid.B106:1315,1991

\bibitem[(Dent 2011)]{Dent}J. B. Dent, S.  Dutta, and E. N. Saridakis, arXiv:1008.1250v1





\bibitem[(Amanullah 2010)]{Aman} Amanullah, R., et al. 2010, Astrophys. J. 716, 712-738
\bibitem[(Reid 2010)]{Reid} Reid, B. A., 2010, Mon. Not. Roy. Astron. Soc. 404, 60-85
\bibitem[(Percival 2010)]{Percival} Percival, W. J., 2010, Mon. Not. Roy. Astron. Soc. 401, 2148-2168
\bibitem[(Wang 2006)]{Wang} Wang, Y. \&  Mukherjee, P. 2006, Astrophys. J. 650, 1
\bibitem[(Bond 1997)]{Bond} Bond, J. R., Efstathiou, G. \&  Tegmark, M. 1997, Mon. Not. Roy. Astron. Soc. 291, L33
\bibitem[(Komatsu 2011)]{Kom} Komatsu, E., et al. 2011, Astrophys. J. Suppl. 192, 18
\bibitem[(Melchiorri 2007)]{Mel} Melchiorri, A., Pagano, L. \&  Pandolfi, S. 2007, Phys. Rev. D 76, 041301
\bibitem[(Komatsu 2009)]{Kom2} Komatsu, E., et al. 2009, Astrophys. J. Suppl. 180:330-376

\end{thebibliography}

\end{document}